\documentclass{article}

\usepackage[english]{babel}
\usepackage[letterpaper,top=2cm,bottom=2cm,left=3cm,right=3cm,marginparwidth=1.75cm]{geometry}
\usepackage{amsmath}
\usepackage[utf8]{inputenc}
\usepackage{graphicx}
\usepackage{subfig}
\usepackage{subcaption}
\usepackage[left]{lineno}
\usepackage{color}
\usepackage{soul}
\usepackage{multirow}
\usepackage{placeins}
\usepackage{booktabs}
\usepackage{bm}
\usepackage{url}

\renewcommand{\arraystretch}{1.2}

\usepackage{xr} 
\title{Simulating the impact of perception bias on social contact surveys for infectious disease modelling}
\author{Thomas J. Harris$^{1,2}$, Prescott C. Alexander$^{2,3}$, Anh B. D. Pham$^{1}$, \\ Joseph Tuccillo$^{4}$, Nicholas Geard\footnote{These authors contributed equally to this paper} $^{1}$, Cameron Zachreson$^{*1}$}
\date{
      $^1$School of Computing and Information Systems, The University of Melbourne\\%
      $^2$Information Systems and Modeling (A-1), Los Alamos National Laboratory\\%
      $^3$Physics and Chemistry of Materials (T-1), Los Alamos National Laboratory\\%
      $^4$Geospatial Science and Human Security Division, Oak Ridge National Laboratory\\[2ex]%
      \today
      }

\begin{document}
\maketitle


\begin{abstract}
Social contact patterns are a key input to many infectious disease models. Contact surveys, where participants are asked to provide information on their recent close and casual contacts with others, are one of the standard methods to measure contact patterns in a population. Surveys that require detailed sociodemographic descriptions of contacts allow for the specification of fine-grained contact rates between subpopulations in models. However, perception biases affecting a surveyed person's ability to estimate sociodemographic attributes (e.g., age, race, socioeconomic status) of others could affect contact rates derived from survey data. Here, we simulate contact surveys using a synthetic contact network of New Mexico to investigate the impact of these biases on survey accuracy and infectious disease model projections. We found that perception biases affecting the estimation of another individual's age and race substantially decreased the accuracy of the derived contact patterns. Using these biased patterns in a Susceptible-Infectious-Recovered compartmental model lead to an underestimation of cumulative incidence among older people (65+ years) and individuals identifying as races other than White. Our study shows that perception biases can impact contact patterns estimated from surveys in ways that systematically underestimate disease burden in minority populations when used in transmission models.
\end{abstract}

\section{Introduction}
Data describing patterns of social contact between subpopulations is a key input for policy-relevant models of infectious disease transmission \cite{wallinga2006using,edmunds1997mixes,davies2020effects,prem2020effect}. Contact patterns within and between sociodemographic groups allow for models that can capture fine-grained epidemic dynamics, otherwise hidden under an assumption of homogenous mixing \cite{manna2024generalized}. As such, there is a growing need to reliably estimate the rate of contact between individuals stratified by multiple sociodemographic attributes \cite{manna2024generalized,di2025individual,manna2024importance,manna2025social,lomas2025modelling,datta2025modelling,ma2021modeling}. These attributes include age, race, ethnicity and socioeconomic status (SES). Models informed by these stratified contact rates can better capture the impact of respiratory virus outbreaks, such as influenza and SARS-CoV-2, which are often felt unequally across sociodemographic subpopulations \cite{irizar2023ethnic,ferguson2022geographic,quinn2011racial,sam2020redefining,bassett2020variation,health_equity_tracker_race,khanijahani2021systematic,harris2025disparities,yellow2020covid}.
\\\\
Contact surveys are a standard method used to measure contact patterns \cite{hoang2019systematic,liu2021rapid,harris2024apparent}. In contact surveys, participants report attributes of their recent close and casual contacts, typically the age of the contact and the interaction setting (e.g., household, workplace, school). To estimate contact rates between subpopulations stratified by multiple sociodemographic attributes, surveys must query additional information beyond the usual age dimension \cite{manna2024generalized}. Correctly identifying different attributes of casual contacts may be difficult because of limited familiarity (e.g., SES of a workplace contact). For these contacts, systematic perception biases may hinder the accurate identification of some attributes \cite{voelkle2012let,feliciano2016shades,kraus2017signs}. For example, targeted studies have shown that Native Americans are racially misidentified at a higher rate than other racial groups in the US \cite{kressin2003agreement}. Because contact surveys rely on the ability of participants to accurately report the attributes of their contacts, biases like these could systematically alter the contact patterns derived from survey data. In this study, we investigate the extent to which perception biases impact the projections of epidemiological models built from biased contact data, and comment on the relevance of the bias impact for public health decision making. 
\\\\
To achieve this, we simulated the collection of contact data through surveys on a high-resolution contact network of New Mexico, USA, built from a census-calibrated synthetic population \cite{tuccillo2023urbanpop,alexander2025epicast}. Within these simulated contact surveys, we incorporated evidence-based perception biases affecting the estimation of contacts' age and race by survey participants. By comparing simulated surveys with varying levels of perception bias to idealised ground truth surveys, we isolated the impact of perception biases on estimated contact rates and outbreak simulations. We found that perception biases affecting the estimation of age and race produced a downward bias in contact rates with older people (65+ years) and individuals identifying as a race other than White. Using these biased contact rates to inform a Susceptible-Infectious-Recovered (SIR) compartmental model of disease spread lead to an underestimation of cumulative incidence in both subpopulations. Our study shows that perception biases can impact contact patterns estimated from surveys in ways that systematically underestimate disease burden in minority populations when used in transmission models. Our results suggest that caution is required when using models calibrated with contact survey data to estimate the impact of infectious disease outbreaks across subpopulations, and highlight a need for further research into how the impact of perception biases can be mitigated.

\section{Methods}
\begin{figure}
    \centering
    \includegraphics[width=\linewidth]{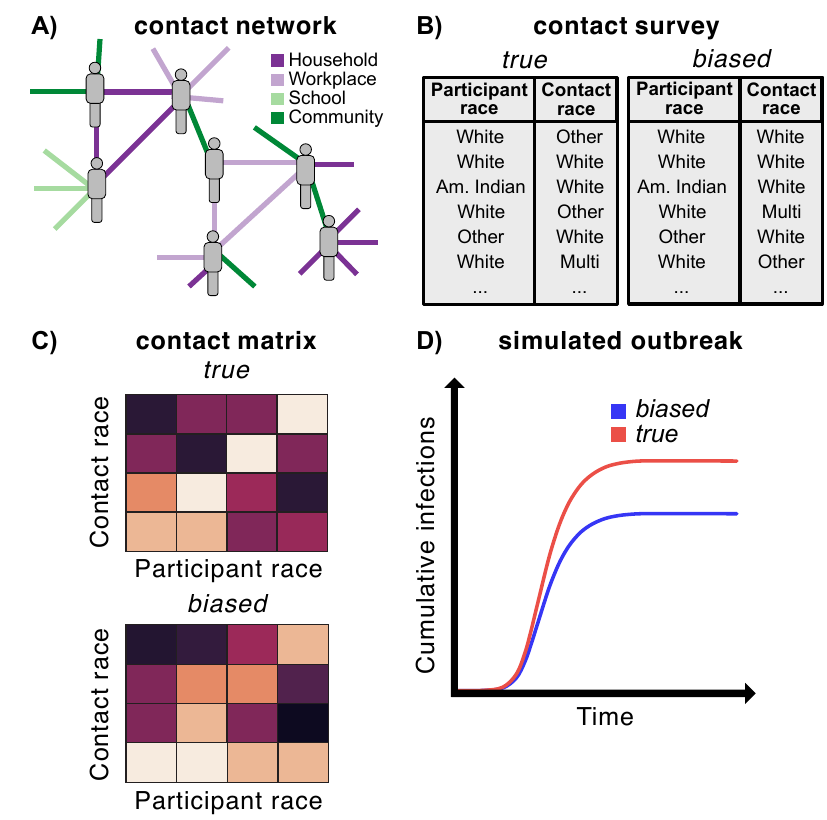}
    \caption{Methodological workflow. We constructed a social contact network (a) from a sociodemographically rich synthetic population of New Mexico, USA. The contact network described interactions between individuals in households, workplaces, schools and community settings. By sampling a set of survey participants from this network, we simulated a contact survey (b) where participants recalled sociodemographic attributes (e.g., race) of their contacts (i.e., graph neighbours) either perfectly (true) or imperfectly due to perception bias (biased). We derived contact matrices (c) from both the true and biased contact data, describing the estimated contact rates within and between sociodemographic groups. To characterise the impact of perception biases on epidemic models, we compared simulated epidemic dynamics (d) using models informed by the true and biased contact data.}
    \label{fig:fig1}
\end{figure}

Our methodological workflow is composed of four key components. First, we sampled contact pairs from a rich synthetic population to form a fine-grained social contact network describing interactions between individuals in households, workplaces, schools and communities (Figure \ref{fig:fig1}A). Second, using this contact network, we simulated a contact survey where participants recalled sociodemographic attributes (e.g., age and race) of their contacts either perfectly (true) or imperfectly (biased) due to simulated perception bias (Figure \ref{fig:fig1}B). We developed a novel probabilistic framework to simulate attribute estimation that allowed us to embed evidence-based perception biases into contact reporting by survey participants. Third, to understand how these perception biases impacted the estimated contact rates within and between sociodemographic groups, we compared contact matrices derived from both the true and biased contact data (Figure \ref{fig:fig1}C). Fourth, to characterise the impact of perception biases on epidemic models, we compared overall and group-specific attack rates simulated using an SIR model informed by the true and biased contact matrices (Figure \ref{fig:fig1}D). Supplementary Material S1 summarises the key model parameters related to each component.
\subsection{Synthetic population}
To produce a realistic synthetic population with detailed individual heterogeneity, we used a synthetic population generation method---UrbanPop \cite{tuccillo2023urbanpop}. UrbanPop is a spatial microsimulation model (SMSM) framework designed to generate synthetic populations with high spatial and sociodemographic resolution. The UrbanPop model draws primarily on data from the American Community Survey (ACS) and the Public Use Microdata Sample (PUMS) provided by the US Census Bureau \cite{census_acs}. We generated a single synthetic population that included individual variables for age, race, ethnicity, household income, and industry of employment (Supplementary Material S2). We categorised race and ethnicity according to the US Office of Management and Budget (OMB) standards used in the 2020 US census. Formally, we described our population of agents as set $A$ with size $|A|=2089388$. Each agent $A_i$, where $i \in {1,2,...,|A|}$, has a set of attributes which we encode symbolically: $\text{age}_i$ describes the age of agent $i$ in years, $\text{race}_i$ describes the racial identity of agent $i$, $\text{ethnicity}_i$ describes the ethnic identity of agent $i$, and $\text{income}_i$ describes the household income stratum of agent $i$. Supplementary Material S3 summarises the synthetic population distribution by age, race, ethnicity and adjusted household income (Supplementary Material S4 outlines household income adjustment method).
\\\\
To simulate daytime contact between individuals in different geographic locations, we extended the approach taken in \cite{tuccillo2023urbanpop, tuccillo_2025_enhancing} by assigning regular daytime locations to agents attending work and school. For agents attending work, this process was based on the US Census Longitudinal Employer-Household Dynamics (LEHD) dataset \cite{census_lodes}, specifically its Origin-Destination Statistics (LODES) product. The LODES data was applied to approximate worker commute flows (i.e., rates of workers moving between residential and employment locations) at the geographic level of census block groups for agents across 13 North American Industry Classification System (NAICS) industry categories. Commute assignments by origin (home) block group were then updated by balancing worker totals with estimated job totals by destination block group and NAICS industry from the LODES Workplace Area Characteristics (WAC) file using Iterative Proportional Fitting (IPF) \cite{lomax2016estimating}. For agents attending school, school commute flows (i.e., rates of school attendees moving between residential and school locations) for students by grade level (spanning primary, secondary, and post-secondary education) were assigned using road network-based proximity of residential street intersections within census block groups to school locations from the Homeland Infrastructure Foundation-Level Data (HIFLD) \cite{hifld}, then also balanced using IPF with official enrolment totals from the National Center for Education Statistics (NCES) \cite{nces}. We applied this procedure across a series of commuting zones overlapping New Mexico, defined by OMB county agglomerations: Combined Statistical Areas (e.g., Albuquerque-Santa Fe-Las Vegas, El Paso-Las Cruces), Metropolitan and Micropolitan Statistical areas (e.g., Farmington, Roswell), and individual ungrouped rural counties.
\\\\
To generate contact networks that simulate setting-specific interactions between individuals, we organised agents into social contact groups based around four main transmission settings: households, schools, workplaces, and communities (split into daytime and nighttime communities). These contact groups describe the sets of individuals that share a physical space in each transmission setting. Each agent can only be a member of one contact group in each transmission setting. Further detail on how contact groups were assigned can be found in Supplementary Material S5.
\\\\
The complete list of agent attributes included in the synthetic population are summarised in Supplementary Material S2.

\subsection{Ground truth contact network}
To construct a social contact network, we sampled contact pairs using a probabilistic process based on shared social contact groups (e.g., households, workplaces, schools). We used these pairs to define an unweighted, undirected graph $F$ which represented the interactions between our set of agents $A$ over the course of one day. Each node in $F$ represented an agent $A_i$. Each edge $E(A_i,A_j)$ indicated at least one interaction had occurred between two individuals $(A_i,A_j)$ where disease transmission could have taken place.
\\\\
We constructed graph $F$ using a probabilistic process, where edges are sampled based on the likelihood of individuals interacting in each transmission setting $k \in \{$household, school, workplace, daytime community, nighttime community$\}$. We defined a probability distribution $P_k(E(A_i,A_j))$ which described the likelihood of an edge $E(A_i,A_j)$ being sampled between any two individuals $(A_i,A_j)$ in transmission setting $k$:
\begin{equation}
\label{edge_sampling}
    P_k(E(A_i,A_j)) = \frac{I_k(A_i,A_j)}{\sum_{g}^{|A|}\sum_{h}^{|A|}I_k(A_g,A_h)}\,. 
\end{equation}
To inform the sampling of edges in $F$, we defined a function $I_k(A_i,A_j)$ specific to transmission setting $k$. This function is equal to 1 if two different individuals $(A_i,A_j)$ share a contact group in transmission setting $k$, and 0 otherwise.
\\\\
To sample edges in the network, we defined a contact rate $c_k$ for each transmission setting $k$, describing the mean number of unique contacts per day for any individual in transmission setting $k$. The overall and setting specific contact rates are outlined in Table \ref{tab:contact rates}. Using these setting-specific contact rates we sampled node pairs without replacement from each of the setting-specific probability distributions $P_{k}$ and assigned edges between these node pairs in our graph $F$. We sampled all edges from each transmission setting in the following order: households, workplaces, schools, nighttime communities, and daytime communities. To ensure we only sampled one edge between two individuals in $F$ (i.e., not allowing parallel edges), we re-normalised each subsequent setting-specific probability distribution based on already sampled edges. Further details on the derivation of transmission setting contact rates and the algorithm for edge sampling can be found in Supplementary Material S6.
\begin{table}[h]
\centering
\begin{tabular}{l|lllll}
\textbf{\bm{$c$}} & \textbf{\bm{$c_\text{H}$}} & \textbf{\bm{$c_\text{S}$}} & \textbf{\bm{$c_\text{W}$}} & \textbf{\bm{$c_\text{DC}$}} & \textbf{\bm{$c_\text{NC}$}} \\ \hline
9.77         & 2.25 (23\%)           & 1.37 (14\%)           & 2.05 (21\%)           & 2.05 (21\%)           & 2.05 (21\%)          
\end{tabular}
\caption{Overall ($c$) and setting-specific ($c_k$) contact rates (contacts per day). Values describe the mean unique contacts per person across a day in households (H), schools (S), workplaces (W) and daytime/nighttime communities (DC/NC). Percentages represent the distribution of overall contact in each setting---derived from \cite{mossong2008social}.}
\label{tab:contact rates}
\end{table}

\noindent

\subsection{Contact survey}
To simulate a contact survey conducted on a given contact network $F$, we sampled a subset of $N_{\text{sample}}$ nodes (i.e., agents) and simulated their individual responses to a contact survey questionnaire. The method for selecting nodes was through uniform random sampling, which produced an approximately representative sample of the population. We defined the set of sampled nodes as ``survey participants", and each node they are connected to as one of their ``contacts". For each simulated contact survey, we assumed participants reported contact behaviour occurring over a single survey day, represented in the synthetic contact network. Surveys that simulate participants' responses unaffected by perception bias are referred to as ``true surveys" and surveys that simulate participants' responses affected by perception bias are referred to as ``biased surveys".

\subsubsection{Survey participant's response}
To simulate a sampled individual $A_i$'s response without perception bias, we iterated through each of their contacts in $F$. For each contact, we recorded the attributes of the survey participant (e.g., $\text{age}_i$), the attributes of their contact (e.g., $\text{race}_j$), and the setting $k$ in which the contact occurred (i.e., transmission setting-specific edge probability distribution $P_k$ from which the edge was sampled).
\\\\
To simulate a sampled individual $A_i$'s response with perception bias, we sampled a value for each contact attribute from a probability distribution. Each distribution described the likelihood of estimating a particular attribute value, given the true attribute value of the contact, and the transmission setting in which the interaction occurred. Below we detail our specific implementation for age and race estimation.

\subsubsection{Age estimation bias ($P_{\text{age}}$)}
To capture variation in the age estimation process for different contacts, we defined a Gamma distribution $P_\text{age}(x)$ from which a survey participant would sample an age estimate. When estimating another individual's age based on visual and audio cues, experimental studies have shown observer accuracy varies with the age of the observed individual \cite{voelkle2012let, norja2022old, moyse2014age}. To simulate this bias in contact surveys, we fit a quadratic polynomial of the mean estimated age $B(\text{age}_j)$, given a contact's age ($\text{age}_j$), using empirical measurements of age estimation bias by subject age from \cite{voelkle2012let} and \cite{norja2022old} (Supplementary Material S7 summarises fitting process). Both studies reflect a consistent experimental finding that the age of older subjects is generally underestimated and, conversely, the age of younger subjects is overestimated \cite{voelkle2012let,norja2022old,vestlund2009experts}. For each reported contact from a participant, we sampled a continuous value from $P_\text{age}(x|\text{age}_j,k,a)$ with a mean and variance defined below:
\begin{equation}
\label{cont_mean}
    \mu= \text{age}_j + \epsilon_{k}a (B(\text{age}_j)-\text{age}_j) \,,
\end{equation}
\begin{equation}
\label{cont_var}
    \sigma^2=\epsilon_{k}h\,,
\end{equation}
where $h$ is a constant describing the maximum distribution variance, and $a$ is a scaling factor affecting the bias magnitude. We defined a parameter $\epsilon_\text{k}$ which captured the impact of the transmission setting on estimation accuracy. The value $\epsilon_\text{k}$ for each transmission setting was estimated from the frequency of exact age recall in each transmission setting, measured in the POLYMOD contact survey \cite{mossong2008social} (Table \ref{tab:setting_scalers}). Survey participants in the POLYMOD study had the option of supplying an exact age estimate for a contact, or an age range if they were uncertain about the contact's age. The frequency of exact age recall in each transmission setting was assumed to be indicative of the familiarity that survey participants had with contacts in that setting. Transmission settings in POLYMOD that were not households, schools or workplaces (e.g., transport, leisure) were classified here as community contacts.
\begin{table}[h]
\centering
\begin{tabular}{llll}
\textbf{\bm{$\epsilon_\text{H}$}} & \textbf{\bm{$\epsilon_\text{S}$}} & \textbf{\bm{$\epsilon_\text{W}$}} & \textbf{\bm{$\epsilon_\text{C}$}} \\ \hline
0           & 0.25           & 0.44             & 0.39          
\end{tabular}
\caption{Setting-specific recall accuracy scaling factors ($\epsilon_k$). Values are derived from the proportion of contacts that participants did not recall the exact age, instead providing an age range, in the POLYMOD contact survey across a day in households (H), schools (S), workplaces (W) and communities (C). Note that while age ranges were provided for a small portion of contacts (9\%) in households (H), we assume household recall is perfectly accurate (i.e., $\epsilon_\text{H}=0$).}
\label{tab:setting_scalers}
\end{table}
\\\\
We varied $a$ between $0\leq a\leq3.87$ to simulate age-related perception biases of varying strength. This range of $a$ was chosen to ensure the mean age estimate $\mu$ was always monotonically increasing with contact age ($\text{age}_j$). When $a=2.56$, the mean age estimate in community contact settings aligned with our experimental fit (i.e., $\mu=B(\text{age}_j)$). We set $h=10$ to encode estimate variation similar to experimental studies of age estimation \cite{norja2022old} (Supplementary Material S7 provides further detail of range and value choice for parameters $a$ and $h$). Finally, as age is typically reported as a whole number, we rounded the sampled value down to the nearest integer value. Table \ref{tab:exp} summarises assumed model settings for investigation of age perception bias.

\subsubsection{Racial estimation bias ($P_{\text{race}}$)}
To simulate the estimation of a contact's racial identity by survey participants, we sampled estimates from a categorical probability distribution $P_{\text{race}}$. Perception of one's own race and the race of others is known to be a complex phenomenon \cite{harris2002multiracial,johnson2023accuracy}. Studies comparing an individual's self-identification of race to an observer's identification in the US have shown that racial misidentification (i.e., when observer-identification does not match self-identification) occurs at significant rates for some racial identities in settings such as healthcare environments \cite{baker2006system,hasnain2004and,kressin2003agreement,kelly1996race,hahn1996identifying,feliciano2016shades}. The observer's attributes, such as their age, race and sex, have been explored as potential reasons for higher rates of misidentification \cite{herman2010you,feliciano2016shades}. Similarly, geographic variation in racial estimation accuracy could reflect familiarity with racial groups common in one's local environment \cite{kelly1996race,harris2002eye}.
\\\\
We defined the probability of sampling a racial category $b$, given the contact's race ($\text{race}_j$), and the transmission setting $k$ and census tract $g$ in which the contact occurred as:
\begin{equation}
\label{p_race}
    P_{\text{race}}(x=b\,|\, \text{race}_j,k,g,r_b)=\begin{cases}
        1-\epsilon_{k} r_b+\zeta_{g,b}\epsilon_{k} r_b &,\,\text{if }b=\text{race}_j \\
        \zeta_{g,b}\epsilon_{k} r_b &,\, \text{otherwise}
    \end{cases}\,,
\end{equation}
where $b \in$ \{White, Black, Asian, American Indian or Alaskan Native (AIAN), Native Hawaiian or Pacific Islander (NHPI), Other, Multiracial\}, $r_b$ is a scaling factor for recall accuracy associated with contacts identifying with race $b$, $\epsilon_\text{k}$ is a scaling factor that captured the impact of the transmission setting $k$ on estimation accuracy (Table \ref{tab:setting_scalers}), and $\zeta_{g,b}$ is the proportion of the population identifying as race $b$ in census tract $g$ where the interaction occurred. In our definition of $P_{\text{race}}$, $1-\epsilon_{k} r_b$ represents the likelihood of a participant correctly identifying a contact's race in transmission setting $k$ based on their true race $b$. The remaining likelihood, $\epsilon_{k} r_b$, is distributed between racial groups based on the racial demographics of the census tract where the interaction occurred $\zeta_{g,b}$.
\\\\
Misidentification of race has often been found to occur at a higher rate among individuals identifying as races other than White (hereafter referred to as non-White) \cite{johnson2023accuracy}. To capture observed differences in the rate of misidentification between non-White individuals and White individuals, we varied the recall accuracy scaling factor for non-White contacts ($r_{NW}$; $r_{\text{b}}=r_{NW}$ for all $b\neq\text{White}$) between 0 and 2.05. This range corresponded with a scaled bias in community settings (i.e., $\epsilon_Cr_{\text{NW}}$) of between 0 and 0.8. We set the recall accuracy scaling factor for White contacts ($r_{\text{W}}$) to 0 (i.e., perfect recall). We assumed attributes of the survey participants did not affect recall (an alternative model where contact recall is dependent on race of the survey participant is described in Supplementary Material S8). Table \ref{tab:exp} summarises assumed model settings for investigation of racial perception bias.
\\\\
Supplementary Material S9 provides a series of visual examples of the simulated estimation process.
\\\\
We described other categorical probability distributions characterising the estimation of contact ethnicity ($P_{\text{ethnicity}}$), where Hispanic individuals are misidentified at a higher rate than non-Hispanic individuals, and income strata ($P_{\text{income}}$), where all income strata are equally likely to be mischaracterised. Description of these distributions and their basis in experimental results can be found in Supplementary Material S10 and S11.

\subsection{Contact matrix}
To derive a contact matrix from simulated contact survey data for a given stratification of the population (e.g., by age group), we computed the mean number of contacts per survey participant reported within and between the population strata. We used the generalised contact matrix definition from \cite{manna2024generalized} to define contact rates between individuals in each population stratum. In this formulation, the group an individual is a member of can be defined as a vector $\textbf{a}=\{\text{age},\text{ race},...\}$. 
\\\\
We defined a generalised contact matrix $G_{\textbf{a},\textbf{b}}$ which describes the mean number of daily contacts reported by a participant in group $\textbf{a}$ with contacts in group $\textbf{b}$:
\begin{equation}
    G_{\textbf{a},\textbf{b}}=\frac{C_{\textbf{a}\rightarrow\textbf{b}}}{N_{\textbf{a}}}
\end{equation}
where $C_{\textbf{a}\rightarrow\textbf{b}}$ is the number of reported contacts between participants in group $\textbf{a}$ and individuals in group $\textbf{b}$, and $N_{\textbf{a}}$ is the total number of survey participants in group $\textbf{a}$.
\\\\
From each contact matrix $G_{\textbf{a},\textbf{b}}$, we computed a contact matrix $G'_{\textbf{a},\textbf{b}}$ that described the per-capita daily contact rate corrected for contact reciprocity between individuals in group $\textbf{a}$ and individuals in group $\textbf{b}$ (further details of these processing steps available in \cite{socialmixr_intro}). 
\\\\
We computed three different types of contact matrices based on one of three sources of contact data: 1) the contacts of all individuals in the synthetic population (ground truth contact matrices), 2) a contact survey where sampled survey participants report all their contacts perfectly (true contact matrices), and 3) a contact survey affected by perception bias where sampled survey participants report all their contacts imperfectly (biased contact matrices).

\subsection{Epidemic simulation}
To understand the practical significance of perception bias in derived contact matrices, we simulated disease spread using a Susceptible-Infectious-Recovered (SIR) compartmental model and compared the epidemic dynamics produced using contact matrices derived from true (i.e., unbiased) survey data to the epidemic dynamics produced using contact matrices derived from biased survey data. We described the rate at which individuals in population group \textbf{a} become infected and recover as below:
\begin{equation}
    \frac{dS_{\textbf{a}}(t)}{dt} = -\beta_{\textbf{a}}S_{\textbf{a}}(t)\sum_{\textbf{b}}G'_{\textbf{a,b}}I_\textbf{b}(t)\,,
\end{equation}
\begin{equation}
    \frac{dI_\textbf{a}(t)}{dt} = \beta_{\textbf{a}}S_{\textbf{a}}(t)\sum_{\textbf{b}}G'_{\textbf{a,b}}I_{\textbf{b}}(t) - \gamma I_{\textbf{a}}(t)\,,
\end{equation}
\begin{equation}
    \frac{dR_{\textbf{a}}(t)}{dt} = \gamma I_{\textbf{a}}(t)\,,
\end{equation}
where $S_\textbf{a}(t)$, $I_\textbf{a}(t)$, and $R_\textbf{a}(t)$ are the populations of susceptible, infected and recovered individuals, respectively, in group $\textbf{a}$ at time $t$; $G'_{\textbf{a,b}}$ is the per-capita contact rate between an individual in group $\textbf{a}$ with individuals in each other population group $\textbf{b}$; $\beta_{\textbf{a}}$ refers to the age-specific susceptibility of individuals in group $\textbf{a}$ to infection; $\gamma$ refers to the recovery rate. We parametrised our model to simulate spread of a COVID-19 like infection (Supplementary Material S12 provides further information on the SIR model design and parametrisation) \cite{billah2020reproductive,davies2020age}.
\\\\
To characterise the impact of perception bias in our epidemic model, we compared group-specific and overall population cumulative infections, disease prevalence and attack rate produced using biased and unbiased survey data. For age-related perception bias, we quantified the impact on older people (65+ years) using an age-stratified SIR model with five year age brackets (Table \ref{tab:exp}). For race-related perception bias, we quantified the impact on non-White individuals using an age and race stratified SIR model with each racial group stratified into five year age brackets (Table \ref{tab:exp}).

\begin{table}[h]
\centering
\resizebox{\columnwidth}{!}{%
\renewcommand{\arraystretch}{1.2}
\begin{tabular}{llll}
\textbf{Attribute} & \textbf{Subpopulation} & \textbf{Perception bias description}                                                                                                                                                                                                                                                                                                                            & \textbf{SIR model}           \\ \hline
Age                                           & Older people (65+)                      & \begin{tabular}[c]{@{}l@{}}Varying age estimation accuracy depending on age of \\contact. Bias magnitude ($a$) varied between \\$0\leq a\leq3.87$.\\Section 2.3.2 outlines estimation process in detail.\end{tabular}                                                                                                                                                                                                                      & Age-structured               \\[1.1cm]
Race                   & non-White                        & \begin{tabular}[c]{@{}l@{}}Varying race estimation accuracy depending on race \\of contact. Racial identification accuracy ($r_{\text{NW}}$) varied \\for non-White contacts $0\leq r_{\text{NW}}\leq2.05$. White contacts\\ assumed to be identified without error ($r_{\text{W}}=0$). Age \\of all contacts estimated perfectly. \\Section 2.3.3 outlines estimation process in detail.\end{tabular} & Age- \& Race-structured      \\[1.7cm]
\end{tabular}
}
\caption{Estimation bias design. For each bias simulation, there is an attribute and subpopulation of interest (`Attribute' \& `Subpopulation'), experimental conditions relating to the estimation bias (`Perception bias description'), and an SIR model stratification approach (`SIR model'). }
\label{tab:exp}
\end{table}

\section{Results} 
\subsection{Estimating ground truth contact matrices}
\begin{figure}
    \centering
    \includegraphics[width=\linewidth]{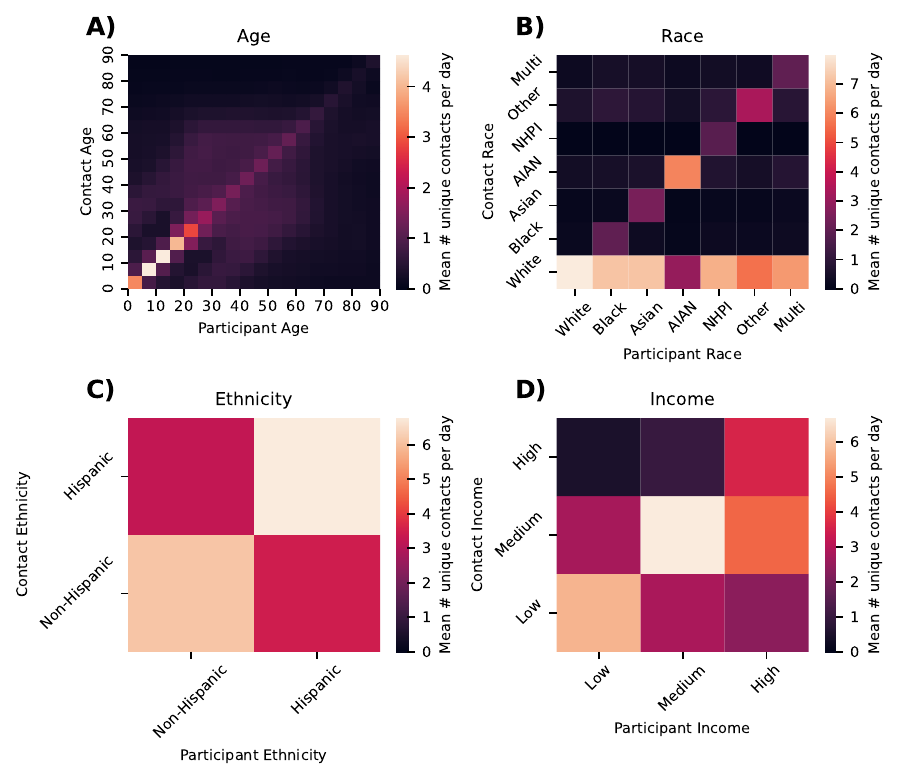}
    \caption{Ground truth contact matrices stratified by age (A), race (B), ethnicity (C) and adjusted household income (D). Our synthetic contact network of the New Mexico population embeds heterogeneity in contact behaviour among individuals. Ground truth matrices are complete representations of the contact patterns in the population (i.e., assuming all individuals are survey participants (x-axis) who report all their contacts (y-axis) perfectly). Each matrix describes the mean number of unique contacts (i.e., graph neighbours) among all individuals in our synthetic contact network, stratified by different sociodemographic attributes. Ground truth contact matrices produced features similar to empirical contact patterns, such as assortative contact by age (A). Note the final age bracket in the age stratified contact matrix (A) represents individuals aged 85+.}
    \label{fig:groundtruth}
\end{figure}

Contact behaviour was assortative (i.e., contact more likely between people with similar characteristics) in ground truth contact matrices computed from the entire contact network, most clearly present in the age-stratified (Figure \ref{fig:groundtruth}A) and ethnicity-stratified (Figure \ref{fig:groundtruth}C) contact patterns. The age-stratified contact patterns showed a tridiagonal structure reflecting inter-generational contact, and elevated contact rates among school age children and working age adults, reflecting school and workplace interactions, respectively (Figure \ref{fig:groundtruth}A). Setting-specific ground truth contact matrices showed varying degrees of assortativity, with household contact patterns being the most assortative across all sociodemographic attributes (Supplementary Material S13). High proportions of some population strata (e.g., White individuals: $\sim75\%$ of total population) produced large reported contact rates relative to other smaller population strata. To account for differences in the size of each population group, we computed the per-capita ground truth contact matrices (Supplementary Material S13).

\subsection{Impact of age-related perception bias}
\begin{figure}
    \centering
    \includegraphics[width=\linewidth]{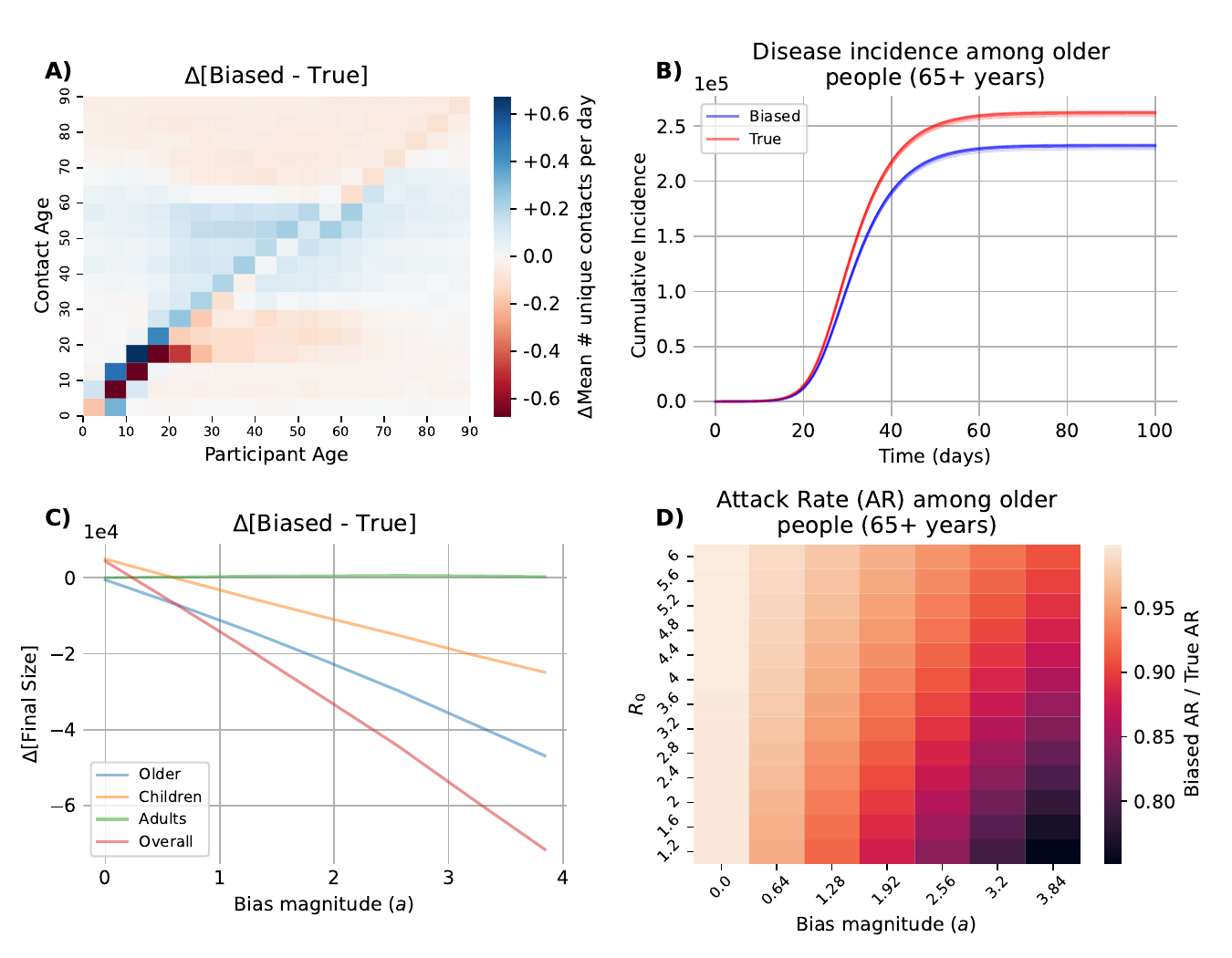}
    \caption{Impact of perception bias affecting the estimation of contact age on estimated contact patterns (A) and epidemic dynamics (B, C \& D). The age of older people (65+ years) is often underestimated by individuals when presented with limited visual and audio stimulus, similar to information available in casual contact settings. By comparing the average contact matrix from ten simulated contact surveys with biased (`Biased') and unbiased (`True') participant recall, surveys affected by age perception bias ($a$) lead to an underestimation of reported contact with older individuals (A). Using an SIR model of disease spread ($R_{0}=2.9$), we found that simulated epidemic dynamics parametrised with the average contact matrix derived from the biased contact data (individual trajectories associated with each simulated contact survey plotted separately with finer line weight) underestimated the true disease burden among people aged 65 years or older (B). Increasing the magnitude of the age estimation bias ($a$) lead to greater underestimation of outbreak final size in children (0-19 years) and older people (65+ years) and a consistent estimation of outbreak final size in the adult population (20-64 years) (C). Differences in subpopulation final size when $a=0$ came about due to simulated variance in the age estimation process even when assuming no directional bias (Equations \ref{cont_mean} \& \ref{cont_var}). The ratio of the simulated attack rate (AR) among older people using the biased contact data and the true contact data (`Biased AR / True AR') decreased when assuming larger $a$ and a lower basic reproduction number (D). Note the final age bracket in the age stratified contact matrix (A) represents individuals aged 85+.}
    \label{fig:fig3}
\end{figure}

We found bias affecting the estimation of contact age ($a=2.56$) lead to a reduction in contact frequency with older individuals (65+ years) (Figure \ref{fig:fig3}A). When comparing the average contact matrix from ten contact surveys (each of sample size $N_{\text{sample}}=10,000$) where we assumed separately biased (`Biased') and unbiased (`True') contact age estimation, participants typically reported between 0.1 and 0.4 fewer daily contacts with older individuals in the biased survey. This reduction reflected an age perception bias where individuals underestimated the age of older contacts.
\\\\
Reduction in the contact rates with older individuals lead to a decrease of $\sim22,000$ cumulative infections among older individuals (Figure \ref{fig:fig3}B; disease prevalence over time provided in Supplementary Material S14). As the magnitude of the age estimation bias ($a$) was increased, the final size of the outbreak decreased (Figure \ref{fig:fig3}C). The outbreak final size among older people (65+ years) and children (0-19 years) decreased, and among adults (20-64 years) remained consistent, reflecting the relative changes in contact rates among these age groups. Similarly, the peak prevalence among older people decreased by $\sim 10,000$ infections (Supplementary Material S14). Differences in subpopulation final size when $a=0$ came about due to simulated variance in the age estimation process even when assuming no directional bias (Equations \ref{cont_mean} \& \ref{cont_var}). 
\\\\
By varying the basic reproduction number $R_0$ (the average number of secondary infections produced by an infected individual in a completely susceptible population) used to calibrate pathogen transmissibility, we observed the ratio of the simulated attack rate (AR) among older people using the biased contact data and the true contact data (`Biased AR / True AR') decreased when assuming larger $a$ and a lower basic reproduction number (Figure \ref{fig:fig3}D). Assuming a lower $R_0=1.4$, where susceptible depletion occurs to a lesser extent, produced a larger proportional difference in the attack rate among older people (Supplementary Material S15).

\subsection{Impact of race-related perception bias}
\begin{figure}
    \centering
    \includegraphics[width=\linewidth]{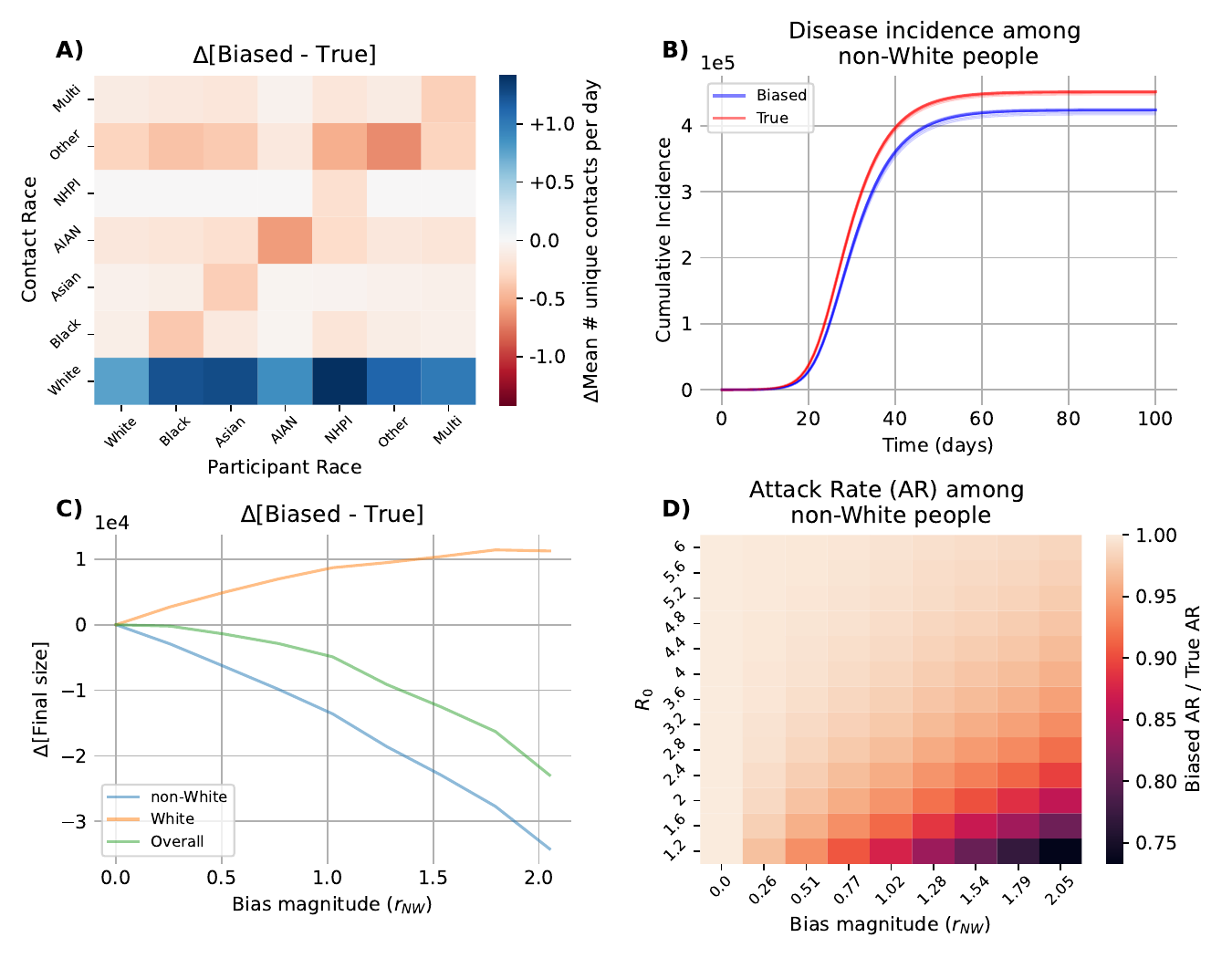}
    \caption{Impact of perception bias affecting the estimation of contact race on estimated contact patterns (A) and epidemic dynamics (B, C \& D). Racial misidentification typically occurs at a higher rate for non-White individuals when observers are presented with limited visual and audio stimulus, similar to information available in casual contact settings. By comparing the average contact matrix from ten simulated surveys with biased (`Biased') and unbiased (`True') recall, racial bias in the identification of non-White individuals ($r_{\text{NW}}=1.79$) lead to a systematic misidentification of non-White individuals as White (A). Using the average contact matrix derived from the biased contact data to simulate disease spread using a SIR model ($R_{0}=2.9$) underestimated the cumulative incidence among non-White individuals (B; individual trajectories associated with each simulated contact survey plotted separately with finer line weight). Increasing the bias magnitude lead to a decrease in outbreak final size among non-White individuals, and an increase in outbreak final size among White individuals (C). The ratio of the simulated attack rate (AR) among non-White individuals using the biased contact data and the true contact data (`Biased AR / True AR') decreased when assuming larger $r_{\text{NW}}$ and a lower basic reproduction number (D).}
    \label{fig:fig4}
\end{figure}

We found bias affecting the estimation of the racial identity of non-White contacts ($r_{\text{NW}}=1.79$) lead to a reduction in contact frequency with non-White individuals (Figure \ref{fig:fig4}A). When comparing the average contact matrix from ten contact surveys (each of sample size $N_{\text{sample}}=10,000$) where we assumed separately biased (`Biased') and unbiased (`True') contact race estimation, participants typically reported between 0.7 and 1.3 fewer daily contacts with non-White individuals in the biased survey results. This reduction reflected a race-related perception bias where non-White contacts are racially misidentified at a higher rate.
\\\\
Reduction in the rates of contact with non-White racial groups lead to a decrease of $\sim30,000$ cumulative infections among non-White individuals (Figure \ref{fig:fig4}B). As the magnitude of race-related perception bias ($r_{\text{NW}}$) was increased, the final size of the outbreak decreased (Figure \ref{fig:fig4}C). The outbreak final size and peak prevalence (Supplementary Material S14) among the non-White population decreased, and among the White population increased, reflecting the relative change in contact rates among racial groups. 
\\\\
By varying the basic reproduction number $R_0$ used to calibrate pathogen transmissibility, we observed the ratio of the simulated attack rate (AR) among non-White individuals using the biased contact data and the true contact data (`Biased AR / True AR') decreased when assuming larger $r_{\text{NW}}$ and a lower basic reproduction number (Figure \ref{fig:fig4}D). Assuming a lower $R_0=1.4$, where susceptible depletion occurs to a lesser extent, produced a larger proportional difference in the attack rate among non-White individuals (Supplementary Material S15).
\\\\
Additional analysis where the SIR model was parametrised without age-specific susceptibility produced a similar reduction in disease spread among older individuals and non-White individuals (Supplementary Material S16).

\section{Discussion}
Contact surveys are widely used as inputs to epidemiological models \cite{liu2021rapid,davies2020effects,prem2020effect}; however, they are subject to biases that may affect model behaviour \cite{hoang2019systematic,gimma2022changes}. In this study, we have characterised a source of bias affecting these surveys stemming from individuals' perception of others. These perception biases introduce systematic reporting inaccuracy in contact data which can in turn degrade estimated contact rates and affect insights derived from epidemic models.
\\\\
Existing contact surveys typically collect data on the age of participants and contacts, which are used to define contact matrices stratified by age \cite{hoang2019systematic}. We found that embedding age-related perception bias in surveys reduced estimated contact rates with older individuals by between 0.1 and 0.4 daily contacts (Figure \ref{fig:fig3}). This perception bias, while well established in experimental studies of age estimation (e.g., US \cite{voelkle2012let}, Finland \cite{norja2022old}, Israel \cite{ganel2022biases}), has not previously been considered in studies of contact behaviour. Of the eight countries surveyed in the POLYMOD contact study, seven countries reported higher per-capita contact rates among older participants (65+ years) with younger people ($<$65 years) than younger participants with older people \cite{mossong2008social,willem2020socrates}. This discrepancy could be an indicator of systematic underestimation of older contacts' age caused by perception bias. To definitively test whether older contacts are being under-reported, specific experimental studies would need to be conducted that simulate survey reporting conditions.
\\\\
When we simulated the impact of perception biases related to the race, ethnicity and income of contacts, we found biases reduced estimated contact rates with specific subpopulations, namely, among non-White individuals, Hispanic individuals and Low income individuals (Figure \ref{fig:fig4}; Supplementary Material S10 and S11). The effect of race-related perception bias on racially stratified contact rates was most notable in the two largest non-White racial groups in New Mexico, namely AIAN and Other. Individuals identifying with either racial group were commonly misidentified as White in our simulated surveys. For AIAN individuals, studies of racial identification report high misidentification rates with individuals often being misidentified as White \cite{campbell2007implications}. For example, 71.21\% of Native Americans in a national study of healthcare settings were incorrectly identified as White \cite{kressin2003agreement}. By comparing the rate of misidentification in our simulated surveys to empirical studies, we found our simulated bias setting ($r_{\text{NW}}=1.79$) aligned with empirical misidentification rates associated with AIAN individuals (Supplementary Material S17 provides visual comparison to empirical rates). Our analysis did not consider the potential effects of participant attributes (e.g., race \cite{feliciano2016shades}) on estimation accuracy. Assuming an alternative model where White survey participants were more likely to classify the race of their contacts as white (i.e., a within-group bias, as observed in \cite{harris2002eye}), produced a similar underestimation of non-White contact rates (Supplementary Material S8).
\\\\
When we used these biased contact rates to simulate disease spread, we found that both the overall outbreak size and attack rate amongst potentially vulnerable groups were underestimated (Figures \ref{fig:fig3} \& \ref{fig:fig4}). Among older people and non-White individuals, we observed a reduction of $\sim 22,000$ and $\sim 30,000$ cumulative infections, respectively. To investigate the source of this underestimation, we conducted a sensitivity analysis on the effect of each transmission setting on outbreak final size, finding that perception bias in community settings was the main driver (Supplementary Material S18). For a given value of $R_0$, differences in final size can occur due to variation in contact assortativity and contact activity level (i.e., the amount of contact) attributed to each population group \cite{manna2024generalized}.
\\\\
Models of infectious disease spread are used to inform public health policy (e.g., hospital resources planning, vaccination strategies, targetted interventions) \cite{hui2021modelling}. Accounting for perception biases in models used to guide public health policy is critical as inaccurate model outputs could lead to poor decision making. For example, modelling that underestimates epidemic disease burden could lead to a subsequent underestimation of required hospital resources. In New Mexico, modelling was used during the COVID-19 pandemic to predict the impact of disease spread on healthcare infrastructure \cite{castro2021new,nm_modelling}. Our simulated underestimation of $\sim10,000$ ($\sim15\%$) peak infections among older people (Supplementary Material S14) could be significant for a disease such as COVID-19 where older people were hospitalised at a disproportionately high rate \cite{health_equity_tracker_age}. Staffed hospital bed capacity in New Mexico was estimated to be $\sim3800$ beds in 2020 \cite{nm_hospital_cap}. Assuming 8.3\% of COVID-19 infections among older people required hospitalisation \cite{AIHW}, a reduction of 10,000 peak infections represents a 21.8\% decrease in predicted hospital bed utilisation due to COVID-19 (Supplementary Material S19). Such an underestimation could further complicate a pandemic response in a state like New Mexico where healthcare access varies among the population \cite{montanez2025medical}.
\\\\
When the allocation of vaccines or the design of targetted interventions is informed by modelling (e.g., \cite{hui2021modelling}), policy decision making could be impacted by underestimates of disease burden in subpopulations. For example, vaccine allocation strategies could be poorly calibrated when prioritising sociodemographic groups (e.g., older people) based on estimated health burden \cite{bullivant2023covid,nomoto2022impact}. Poorly calibrated allocation strategies could lead to excess hospitalisations and deaths due to fewer vaccines being made available to those groups who are most at risk. Similarly, inaccurate estimates could impact the design of targeted interventions. In response to the elevated COVID-19 risk experienced by Indigenous Peoples, specific pandemic policy was implemented in First Nations communities to reduce transmission \cite{power2020covid}. For example, movement controls in remote Aboriginal communities in Australia, and mandatory masking in public spaces in Navajo Nation in the US. Design of these policies must factor in the estimated risk posed by disease transmission alongside the social and economic burden transmission controls place on communities \cite{power2020covid}. Policy design that is informed by inaccurate model estimates will poorly account for the true disease burden likely to impact these communities.
\\\\
A strength of this study is the realistic contact network underpinning our survey simulation. Contact networks have been used to characterise different aspects of social mixing relevant to epidemic spread \cite{eubank2006structure, glass2008social,zhang2015modeling,barrett2009generation}. In order to simulate detailed contact surveys, we required a contact network with fine-grained heterogeneity in individual attributes and setting-specific interactions. We built upon recent work simulating respiratory pathogen spread in large-scale high resolution synthetic populations to construct a realistic social contact network for simulating contact surveys \cite{tuccillo2023urbanpop,alexander2025epicast}. 
\\\\
The probability of two individuals interacting in our network was driven by social contact groups that described sets of individuals that share a physical space in specific transmission settings (e.g. households). While empirical contact data was not used to inform the probability of interaction, the patterns of contact behaviour in our network aligned with existing experimental measurements. For example, the age-stratified contact patterns in our contact network (Figure \ref{fig:groundtruth}A) exhibited three common features found in empirical studies: 1) assortative contact by age, 2) a tridiagonal structure reflecting inter-generational contact (driven by household contact), and 3) elevated contact rates among school age children and working age adults, reflecting school and workplace interactions, respectively \cite{mossong2008social,gimma2022changes}. Similarly, empirical estimates of contact rates in Hungary and Switzerland revealed assortative contact by SES which aligned with the contact assortativity by household income identified here (Figure \ref{fig:groundtruth}D) \cite{manna2024generalized,koltai2022reconstructing,di2025individual}. When estimating contact patterns by race, we saw a heightened racial contact assortativity among AIAN individuals (Figure \ref{fig:groundtruth}B). This feature (most distinct in the community, workplace and school settings) is reflective of regions with large AIAN populations, such as the Tribal Lands associated with the 23 Native Nations in New Mexico \cite{yellow2020covid}.
\\\\
This study is one of the first to characterise contact patterns at such high resolution driven only by structural patterns in a census-calibrated synthetic population. The UrbanPop framework captures observed correlations between sociodemographic attributes and attributes used in the assignment of contact groups which drive contact behaviour in our network model \cite{tuccillo2023urbanpop,alexander2025epicast}. For example, we observed assortative contact by ethnicity in workplaces (Supplementary Material S13) which came about due to ethnic correlations in employment industry and work location, both of which are used to assign work groups in our network model (Supplementary Material S5) \cite{tuccillo2023urbanpop,harris2025disparities}. Notably, no assumption of assortativity within contact groups was required to produce patterns resembling empirical data. We assumed contact groups were perfectly mixed (i.e., the likelihood of any two individuals contacting one another within each group was equal (Equation \ref{edge_sampling})). The similarity of our ground truth contact matrices to empirical data could suggest contact assortativity in social networks is more strongly influenced by the homogeneity of attributes in contact groups (i.e., induced homophily) and less so by individual assortative contact selection within contact groups (i.e., choice homophily) \cite{au2023theoretical,kossinets2009origins}. Future work could include a more rigorous comparison of our simulated network to empirical contact data and characterisation of contact networks under different assumptions of assortativity within contact groups.

\subsection{Limitations}
Our representation of contact attribute estimation approximates a complex human process. We derived our estimation model from both qualitative and quantitative findings in experimental studies of attribute estimation (Supplementary Material S7 \& S17). In these studies, observers were typically asked to estimate subject attributes based on static stimuli (e.g., photographs \cite{voelkle2012let,norja2022old}). In reality, survey participants may use information such as a contact's body movements, vocal patterns or name (if known) to inform estimation \cite{hilliar2008barack,feliciano2016shades}. Future work could include analysis of estimation models that factor in a broader set of inputs likely available to survey participants.
\\\\
Furthermore, we assumed the estimation of different attributes in contacts were independent processes. There is evidence that these processes intersect. For example, in a Belgian study of racial identification, Caucasian observers performed better at identifying Caucasian subject's ages, than African subject's ages \cite{dehon2001other}. We note that rules that include both the subject and observer's age and race in the age estimation process could be easily described using our model framework. We leave investigation of such multi-attribute rules to future work.
\\\\
We identify four further limitations in the design of our study. First, the setting-specific recall accuracy scaling factors ($\epsilon_k$) were derived from only contact age estimates in the POLYMOD contact study (Supplementary Material S6) \cite{mossong2008social}. These scaling factors were assumed to reflect uncertainty when estimating age and race in different transmission settings. Second, contact under-reporting or over-reporting, where contacts are incorrectly left out or included in responses, was not considered in this study. Given the increased burden on survey participants in a detailed sociodemographic survey, misreporting of contacts by participants is possible. Third, the relative contact rate in each transmission setting ($c_k$) was derived from the POLYMOD contact study, a set of large contact surveys conducted in Europe. Measurements of setting-specific contact rates in the US (e.g., \cite{taube2025characterising}) could be used to assess the robustness of our findings. Finally, we used the racial demographics of each census tract to inform participants' estimation of contact race. Future work could assess the robustness of our findings under alternative geographic stratifications (e.g., census block group, county).

\subsection{Conclusion}
We characterised the impact of age- and race-related perception biases on contact matrices and epidemic models parametrised from contact surveys. For both perception biases, we found contact rates associated with a minority group were reduced and, subsequently, disease burden associated with this group was underestimated. While the mechanisms driving these perception biases are not well understood, there is substantial evidence that familiarity with a sociodemographic attribute dictates the likelihood of correctly identifying the attribute in others \cite{feliciano2016shades,harris2002eye,voelkle2012let}. If this is a general principle, the disease burden in minority groups will typically be underestimated when relying on contact survey data. As minority groups often represent some of the most vulnerable subpopulations, these perception biases could lead to modelling that poorly serves the communities most at risk in a pandemic. Our results suggest that caution is required when using such modelling to inform public health decision making, and highlight the need for further research into how the impact of perception biases can be mitigated.

\section{Acknowledgments}
We would like to thank Dr Rob Moss, Melbourne School of Population and Global Health, for helpful discussions during the course of this study.
\\\\
Thomas Harris is supported by an Australian Government Research Training Program Scholarship and The University of Melbourne Elizabeth and Vernon Puzey Scholarship. 
\\\\
Prescott C. Alexander was supported by the US Department of Energy, Office of Science, Office of Advanced Scientific Computing Research and by cooperative agreement CDC-RFA-FT-23-0069 from the CDC’s Center for Forecasting and Outbreak Analytics. The contents of this paper are solely the responsibility of the authors and do not necessarily represent the official views of the Centers for Disease Control and Prevention.
\\\\
This work is approved by Los Alamos National Laboratory (LANL) for public distribution under LA-UR-25-29885. The findings and conclusions in this report are those of the authors and do not necessarily represent the official position of LANL.

\section{Author Contributions}

T.J.H.: conceptualization, data curation, formal analysis, investigation, methodology, software, validation, visualization, writing – original draft, writing – review \& editing. P.C.A.: data curation, methodology, software, writing – review \& editing. A.B.D.P.: conceptualization, writing – review \& editing. J.T.: data curation, resources, writing – review \& editing. N.G.: conceptualization, investigation, methodology, supervision, writing – review \& editing. C.Z.: conceptualization, investigation, methodology, supervision, writing – review \& editing.

\section{Data Accessibility}

The data and code associated with this manuscript can be accessed on GitHub at \url{https://github.com/tomharris4/contact-survey-simulation}. 

\newpage
\bibliographystyle{vancouver}
\bibliography{ref}

\end{document}


\newpage

\newcommand{\beginsupplement}{%

 \setcounter{table}{0}
   \renewcommand{\thetable}{S\arabic{table}}%
   
     \setcounter{figure}{0}
      \renewcommand{\thefigure}{S\arabic{figure}}%
      
      \setcounter{page}{1}
      \renewcommand{\thepage}{S\arabic{page}} 
      
      \setcounter{section}{0}
      \renewcommand{\thesection}{S\arabic{section}}
      
      \setcounter{equation}{0}
      \renewcommand{\theequation}{S\arabic{equation}}
     }

\beginsupplement

\FloatBarrier
{\bf \Large{Supporting Information for:\\ Simulating the impact of perception bias on detailed sociodemographic surveys of social contact behaviour}}

\section{Model parameter summary}
\label{model_params}
\begin{table}[h]
\centering
\resizebox{\textwidth}{!}{%
\begin{tabular}{|l|l|l|l|}
\hline
\multicolumn{1}{|l|}{\textbf{Model component}} & \textbf{Parameter} & \multicolumn{1}{l|}{\textbf{Description}} & \multicolumn{1}{l|}{\textbf{Data sources}} \\ \hline
\multicolumn{1}{|l|}{Synthetic population} & $A_i \in A$ & \multicolumn{1}{l|}{\begin{tabular}[c]{@{}l@{}}Set of agents described in synthetic \\ population ($|A|=2089388$). Each \\ agent $A_i$, where $i \in {1,2,...,|A|}$, \\ has a set of attributes which we encode\\ symbolically.\end{tabular}} & \multicolumn{1}{l|}{UrbanPop \cite{tuccillo2023urbanpop}} \\ \cline{2-4} 
\multicolumn{1}{|l|}{} & $\text{age}_i$ & \multicolumn{1}{l|}{Age (in years) of agent $A_i$.} & \multicolumn{1}{l|}{UrbanPop \cite{tuccillo2023urbanpop}} \\ \cline{2-4} 
\multicolumn{1}{|l|}{} & $\text{race}_i$ & \multicolumn{1}{l|}{\begin{tabular}[c]{@{}l@{}}Race (White, Black, Asian, \\ American Indian or Alaskan Native \\ (AIAN), Native Hawaiian or Pacific \\ Islander (NHPI), Other, Multiple) of \\ agent $A_i$.\end{tabular}} & \multicolumn{1}{l|}{UrbanPop \cite{tuccillo2023urbanpop}} \\ \cline{2-4} 
\multicolumn{1}{|l|}{} & $\text{ethnicity}_i$ & \multicolumn{1}{l|}{\begin{tabular}[c]{@{}l@{}}Ethnicity (Hispanic, non-Hispanic) \\ of agent $A_i$.\end{tabular}} & \multicolumn{1}{l|}{UrbanPop \cite{tuccillo2023urbanpop}} \\ \cline{2-4} 
\multicolumn{1}{|l|}{} & $\text{income}_i$ & \multicolumn{1}{l|}{\begin{tabular}[c]{@{}l@{}}Adjusted household income (Low, \\ Medium, High) of agent $A_i$. \\ Income adjusted for household size.\\ Strata determined using method \\ based on US national median.\\ See Supplementary Material \\ \ref{hh_income_method} for details.\end{tabular}} & \multicolumn{1}{l|}{UrbanPop \cite{tuccillo2023urbanpop}} \\ \hline
\end{tabular}}
\caption{}
\end{table}
\clearpage
\begin{table}[]
\centering
\resizebox{\textwidth}{!}{%
\begin{tabular}{|l|l|l|l|}
\hline
\multicolumn{1}{|l|}{\textbf{Model component}} & \textbf{Parameter} & \multicolumn{1}{l|}{\textbf{Description}} & \multicolumn{1}{l|}{\textbf{Data sources}} \\ \hline
\multicolumn{1}{|l|}{Contact network} & $F$ & \multicolumn{1}{l|}{\begin{tabular}[c]{@{}l@{}}Unweighted, undirected graph without \\ parallel edges describing interactions \\ between agents over the course of one \\ day. Each node represented an agent \\ $A_i$. Each edge $E(A_i,A_j)$ \\ indicated at least one interaction had \\ occurred between two individuals.\end{tabular}} & \multicolumn{1}{l|}{-} \\ \cline{2-4} 
\multicolumn{1}{|l|}{} & $P_k(E(A_i,A_j))$ & \multicolumn{1}{l|}{\begin{tabular}[c]{@{}l@{}}Probability distribution of an edge \\ $E(A_i,A_j)$ being sampled between \\ any two individuals $(A_i,A_j)$ in \\ transmission setting $k$.\end{tabular}} & \multicolumn{1}{l|}{-} \\ \cline{2-4} 
\multicolumn{1}{|l|}{} & $I_k(A_i,A_j)$ & \multicolumn{1}{l|}{\begin{tabular}[c]{@{}l@{}}Interaction function equal to 1 \\ if two different individuals $(A_i,A_j)$ \\ share a contact group in transmission \\ setting $k$, otherwise 0. \\ Used in derivation of $P_k(E(A_i,A_j))$.\end{tabular}} & \multicolumn{1}{l|}{-} \\ \cline{2-4}  
\multicolumn{1}{|l|}{} & $c$ & \multicolumn{1}{l|}{\begin{tabular}[c]{@{}l@{}}Overall contact rate. Mean number \\ of unique contacts per day in any \\ transmission setting. $c=9.77$ \\ contacts per day.\end{tabular}} & \multicolumn{1}{l|}{POLYMOD \cite{mossong2008social}} \\ \cline{2-4} 
\multicolumn{1}{|l|}{} & $c_H$ & \multicolumn{1}{l|}{\begin{tabular}[c]{@{}l@{}}Household contact rate. Mean \\ number of unique contacts per \\ day in household setting. \\ $c_H=2.25$ contacts per day.\end{tabular}} & \multicolumn{1}{l|}{POLYMOD \cite{mossong2008social}} \\ \cline{2-4} 
\multicolumn{1}{|l|}{} & $c_S$ & \multicolumn{1}{l|}{\begin{tabular}[c]{@{}l@{}}School contact rate. Mean \\ number of unique contacts per \\ day in school setting. \\ $c_S=1.37$ contacts per day.\end{tabular}} & \multicolumn{1}{l|}{POLYMOD \cite{mossong2008social}} \\ \cline{2-4} 
\multicolumn{1}{|l|}{} & $c_W$ & \multicolumn{1}{l|}{\begin{tabular}[c]{@{}l@{}}Workplace contact rate. Mean \\ number of unique contacts per \\ day in workplace setting. \\ $c_W=2.05$ contacts per day.\end{tabular}} & \multicolumn{1}{l|}{POLYMOD \cite{mossong2008social}} \\ \cline{2-4} 
\multicolumn{1}{|l|}{} & $c_{DC}$, $c_{NC}$ & \multicolumn{1}{l|}{\begin{tabular}[c]{@{}l@{}}Daytime/Nighttime community \\ contact rate. Mean number of \\ unique contacts per day in \\ community setting. \\ $c_{DC}=c_{NC}=2.05$ contacts per day.\end{tabular}} & \multicolumn{1}{l|}{POLYMOD \cite{mossong2008social}} \\ \hline
\end{tabular}}
\caption{}
\end{table}
\clearpage
\begin{table}[]
\centering
\resizebox{\textwidth}{!}{%
\begin{tabular}{|l|l|l|l|}
\hline
\multicolumn{1}{|l|}{\textbf{Model component}} & \textbf{Parameter} & \multicolumn{1}{l|}{\textbf{Description}} & \multicolumn{1}{l|}{\textbf{Data sources}} \\ \hline
\multicolumn{1}{|l|}{Contact survey} & $N_{\text{sample}}$ & \multicolumn{1}{l|}{\begin{tabular}[c]{@{}l@{}}Survey sample size (i.e., number \\ of participants).\end{tabular}} & \multicolumn{1}{l|}{-} \\ \cline{2-4} 
\multicolumn{1}{|l|}{} & $P_{\text{age}}$ & \multicolumn{1}{l|}{\begin{tabular}[c]{@{}l@{}}Gamma distribution from \\ which a survey participant would \\ sample a contact age estimate.\end{tabular}} & \multicolumn{1}{l|}{-} \\ \cline{2-4} 
\multicolumn{1}{|l|}{} & $B(\text{age}_j)$ & \multicolumn{1}{l|}{\begin{tabular}[c]{@{}l@{}}Function mapping a contact's \\ age to empirical bias fit.\end{tabular}} & \multicolumn{1}{l|}{\begin{tabular}[c]{@{}l@{}}Voelke et al. \cite{voelkle2012let} \\ Norja et al. \cite{norja2022old}\end{tabular}} \\ \cline{2-4} 
\multicolumn{1}{|l|}{} & $a$ & \multicolumn{1}{l|}{\begin{tabular}[c]{@{}l@{}}Scaling factor affecting the bias \\ magnitude. Default value $a=2.56$ \\ (aligns with empirical estimates). \\ Varied between 0 and 3.87 in \\ main analysis.\end{tabular}} & \multicolumn{1}{l|}{-} \\ \cline{2-4} 
\multicolumn{1}{|l|}{} & $\epsilon_\text{H}$ & \multicolumn{1}{l|}{\begin{tabular}[c]{@{}l@{}}Household-specific scaling factor \\ for estimation accuracy. \\ $\epsilon_\text{H}=0$. Note \\ that while age ranges were provided \\ for a small portion of contacts (9\%) \\ in households, we assume \\ household recall is perfectly accurate.\end{tabular}} & \multicolumn{1}{l|}{POLYMOD \cite{mossong2008social}} \\ \cline{2-4} 
\multicolumn{1}{|l|}{} & $\epsilon_\text{W}$ & \multicolumn{1}{l|}{\begin{tabular}[c]{@{}l@{}}Workplace-specific scaling factor \\ for estimation accuracy. \\ $\epsilon_\text{H}=0.42$.\end{tabular}} & \multicolumn{1}{l|}{POLYMOD \cite{mossong2008social}} \\ \cline{2-4} 
\multicolumn{1}{|l|}{} & $\epsilon_\text{S}$ & \multicolumn{1}{l|}{\begin{tabular}[c]{@{}l@{}}School-specific scaling factor \\ for estimation accuracy. \\ $\epsilon_\text{S}=0.25$.\end{tabular}} & \multicolumn{1}{l|}{POLYMOD \cite{mossong2008social}} \\ \cline{2-4} 
\multicolumn{1}{|l|}{} & $\epsilon_\text{NC}$, $\epsilon_\text{DC}$ & \multicolumn{1}{l|}{\begin{tabular}[c]{@{}l@{}}Nighttime/Daytime \\ community-specific scaling factor \\ for estimation accuracy. \\ $\epsilon_\text{NC}=\epsilon_\text{DC}=0.39$.\end{tabular}} & \multicolumn{1}{l|}{POLYMOD \cite{mossong2008social}} \\ \cline{2-4} 
\multicolumn{1}{|l|}{} & $h$ & \multicolumn{1}{l|}{\begin{tabular}[c]{@{}l@{}}Constant describing the maximum \\ age estimation distribution variance.\\ $h=10$.\end{tabular}} & \multicolumn{1}{l|}{-} \\ \hline
\end{tabular}}
\caption{}
\end{table}
\clearpage
\begin{table}[]
\centering
\resizebox{\textwidth}{!}{%
\begin{tabular}{|l|l|l|l|}
\hline
\multicolumn{1}{|l|}{\textbf{Model component}} & \textbf{Parameter} & \multicolumn{1}{l|}{\textbf{Description}} & \multicolumn{1}{l|}{\textbf{Data sources}} \\ \hline 
\multicolumn{1}{|l|}{Contact survey} & $P_{\text{race}}$ & \begin{tabular}[c]{@{}l@{}}Categorical probability distribution \\ for sampling a racial category \\ describing a contact $b \in$ \{White, \\ Black, Asian, American Indian or \\ Alaskan Native (AIAN), Native \\ Hawaiian or Pacific Islander \\ (NHPI), Other, Multiple\}.\end{tabular} & - \\ \cline{2-4} 
& $r_b$ & \begin{tabular}[c]{@{}l@{}}Scaling factor for recall accuracy \\ associated with contacts identifying \\ with race $b$. For non-White \\ contacts ($r_\text{b}=r_{NW}$ for all $b\neq\text{White}$),\\ scaling factor varied between 0 \\ and 2.05 in main analysis. For \\ White contacts ($r_{\text{W}}$), \\ scaling factor set to 0.\end{tabular} & \begin{tabular}[c]{@{}l@{}}Kressin et al. \cite{kressin2003agreement}, \\ Campbell et al. \cite{campbell2007implications},\\ Kelly et al. \cite{kelly1996race}, \\ Gomez et al. \cite{gomez2006misclassification}\end{tabular} \\ \cline{2-4} 
& $\zeta_{g,b}$ & \begin{tabular}[c]{@{}l@{}}Proportion of the population \\ identifying as race $b$ in \\ census tract $g$ where \\ the interaction occurred.\end{tabular} & UrbanPop \cite{tuccillo2023urbanpop} \\ \hline
\end{tabular}}
\caption{}
\end{table}
\clearpage
\begin{table}[]
\centering
\resizebox{\textwidth}{!}{%
\begin{tabular}{|l|l|l|l|}
\hline
\multicolumn{1}{|l|}{\textbf{Model component}} & \textbf{Parameter} & \multicolumn{1}{l|}{\textbf{Description}} & \multicolumn{1}{l|}{\textbf{Data sources}} \\ \hline
\multicolumn{1}{|l|}{Contact matrix} & $G_{\textbf{a},\textbf{b}}$ & \begin{tabular}[c]{@{}l@{}}Mean number of daily contacts \\ reported by a participant in \\ group $\textbf{a}$ with \\ contacts in group $\textbf{b}$\end{tabular} & - \\ \cline{2-4} 
& $C_{\textbf{a}\rightarrow\textbf{b}}$ & \begin{tabular}[c]{@{}l@{}}Total number of reported contacts \\ between participants in group \\ $\textbf{a}$ and individuals in \\ group $\textbf{b}$.\end{tabular} & - \\ \cline{2-4} 
& $N_{\textbf{a}}$ & \begin{tabular}[c]{@{}l@{}}Total number of survey \\ participants in group $\textbf{a}$.\end{tabular} & - \\ \cline{2-4} 
& $G'_{\textbf{a},\textbf{b}}$ & \begin{tabular}[c]{@{}l@{}}Per-capita contact rate corrected \\ for contact reciprocity\\ between individuals in \\ group $\textbf{a}$ and \\ individuals in group $\textbf{b}$.\end{tabular} & - \\ \hline
\end{tabular}}
\caption{}
\end{table}
\clearpage
\begin{table}[]
\centering
\resizebox{\textwidth}{!}{%
\begin{tabular}{|l|l|l|l|}
\hline
\multicolumn{1}{|l|}{\textbf{Model component}} & \textbf{Parameter} & \multicolumn{1}{l|}{\textbf{Description}} & \multicolumn{1}{l|}{\textbf{Data sources}} \\ \hline
\multicolumn{1}{|l|}{Epidemic simulation} & $S_\textbf{a}(t)$, $I_\textbf{a}(t)$, $R_\textbf{a}(t)$ & \begin{tabular}[c]{@{}l@{}}Populations of susceptible, infected \\ and recovered individuals, respectively, \\ in group $\textbf{a}$ at some time $t$.\end{tabular} & - \\ \cline{2-4} 
& $\beta_{\textbf{a}}$ & \begin{tabular}[c]{@{}l@{}}Age-specific susceptibility of \\ individuals in group $\textbf{a}$ \\ to infection. See Supplementary \\ Material \ref{ngm}.\end{tabular} & Davies et al. \cite{davies2020age}\\ \cline{2-4} 
& $\gamma$ & \begin{tabular}[c]{@{}l@{}}Recovery rate. See Supplementary \\ Material \ref{ngm}.\end{tabular} & - \\ \hline
\end{tabular}}
\caption{}
\end{table}

\clearpage
\section{Synthetic population agent attributes}
\label{agent_list}
\begin{table}[h]
\begin{tabular}{|l|l|}
\hline
\textbf{Attribute}        & \textbf{Description}                                                                                                                                                \\ \hline
Age                       & discrete - years  (UrbanPop)                                                                                                                                                  \\ \hline
Race                      & \begin{tabular}[c]{@{}l@{}}categorical - \{White, Black, Asian, American Indian or Alaskan Native, \\ Native Hawaiian or Pacific Islander, Other, Multiple\} (UrbanPop)\end{tabular} \\ \hline
Ethnicity                 & categorical - \{Non-Hispanic, Hispanic\} (UrbanPop)                                                                                                                            \\ \hline
Household income          & discrete - dollars (UrbanPop)                                                                                                                                                  \\ \hline
Household group           & string identifier - shared between household members (UrbanPop)                                                                                                                \\ \hline
School                    & string identifier - shared between agents attending same school (UrbanPop)                                                                                                     \\ \hline
School grade              & discrete - \{1, 2, 3,..., 14, 15\} (UrbanPop)                                                                                                                                  \\ \hline
School group              & string identifier - shared between school group members                                                                                                             \\ \hline
Industry of Employment    & categorical - three digit NAICS code \{111-999\} (UrbanPop)                                                                                                                    \\ \hline
Work group                & string identifier - shared between work group members                                                                                                               \\ \hline
Daytime community group   & \begin{tabular}[c]{@{}l@{}}string identifier - census blockgroup attended during the daytime; \\ shared between daytime community members (UrbanPop)\end{tabular}              \\ \hline
Nighttime community group & \begin{tabular}[c]{@{}l@{}}string identifier - census blockgroup attended during the nighttime; \\ shared between nighttime community members (UrbanPop)\end{tabular}          \\ \hline
\end{tabular}
\caption{Agent attributes. For each attribute, the data type is provided, along with a description of the attribute meaning and possible attribute values. Attributes included in the UrbanPop synthetic population are marked with `(UrbanPop)'}
\label{tab:agent_list}
\end{table}

\clearpage
\section{New Mexico synthetic population demographics}
\label{pop_dist}
\begin{figure}[h]
    \centering
    \includegraphics[width=0.7\linewidth]{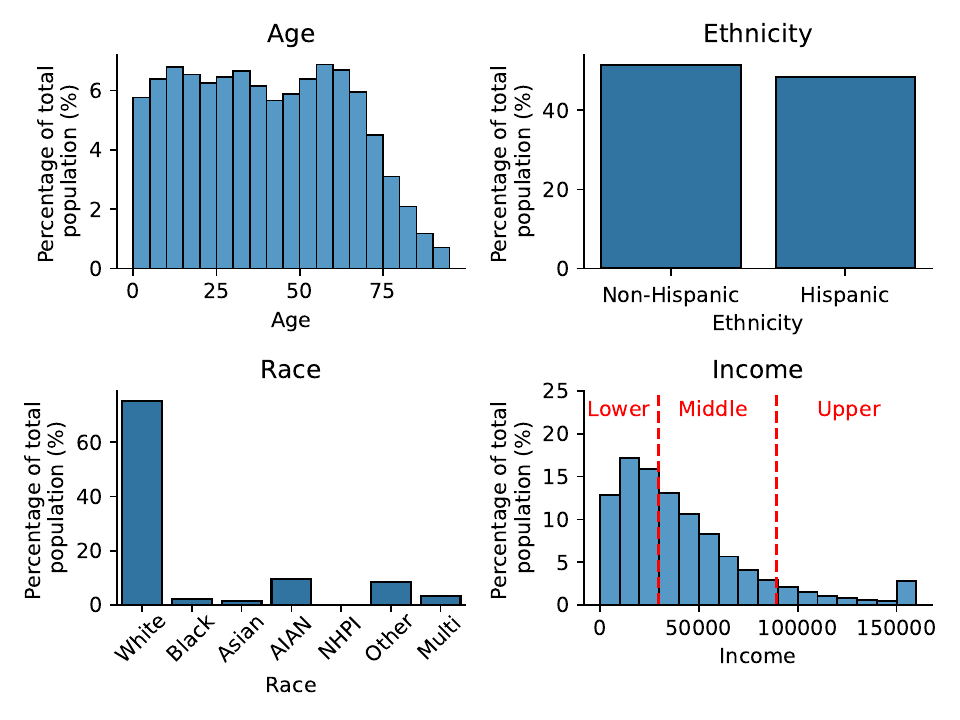}
    \caption{Demographic characteristics of the New Mexico synthetic population. Percentage of the population in each five-year age bracket (top left), ethnic category (top right), racial category (bottom left) and adjusted household income (USD) bracket (bottom right). Note the final brackets in age and income distributions represent individuals aged 90+ and earning \$150000+, respectively.}
    \label{fig:supp_demo}
\end{figure}

\begin{figure}[h]
    \centering
    \includegraphics[width=1\linewidth]{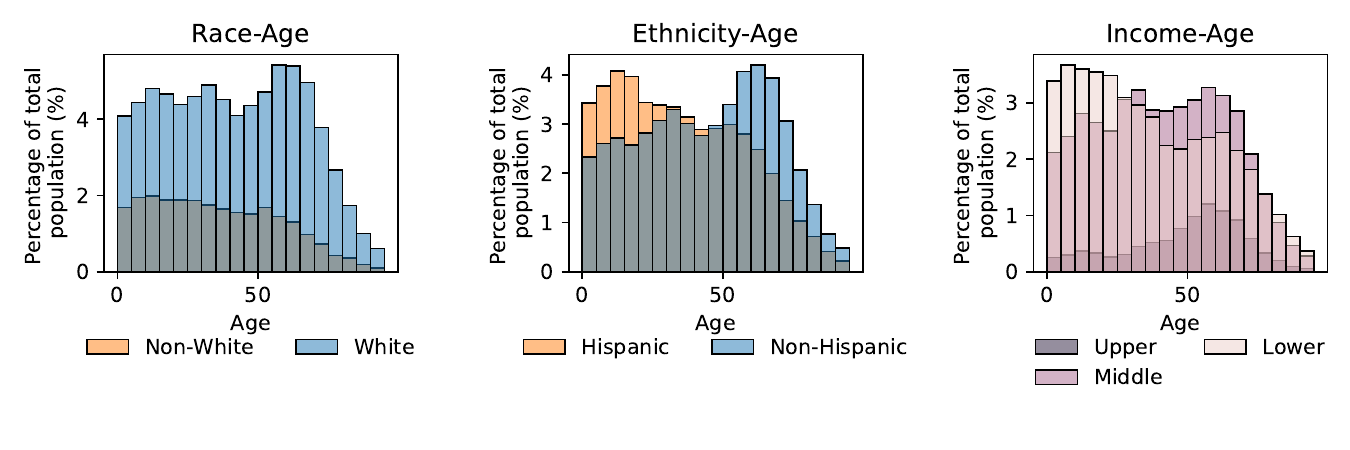}
    \caption{Age distribution of the New Mexico synthetic population by race (left), ethnicity (centre) and adjusted household income (right). Note the final bracket in each age distribution represents individuals aged 90+.}
    \label{fig:supp_demo_age}
\end{figure}

We summarised the main demographic characteristics of the synthetic population informing our social contact network. Namely, the distribution of age, race, ethnicity and adjusted household income (Figure \ref{fig:supp_demo}). Given the inclusion of age in all our stratified SIR model formulations, we also provide the difference in age distribution by race, ethnicity and income (Figure \ref{fig:supp_demo_age}).

\clearpage
\section{Household income adjustment method}
\label{hh_income_method}
We adjusted household income to account for differences in expenditure related to household size using the square root equivalence scale \cite{dudel_assessing_2021}. Using this approach, we divided household income by the square root of household size to produce an adjusted household income:
\begin{equation}
    \text{Adjusted Household Income}=\frac{\text{Household Income}}{\sqrt{\text{Household Size}}} \, .
\end{equation}
We grouped individuals into three income strata: low income, medium income and high income. To define an individual as low, medium or high, we used an existing definition of social class where middle class is defined as having an adjusted household income that is between two thirds and double the US national median adjusted household income \cite{pew_class,census_hh_income}. Anyone with an adjusted household income below two thirds is classified as low income, and anyone with an adjusted income greater than double is classified as high income. Supplementary Material \ref{pop_dist} summarises the state population distribution of adjusted household income.

\clearpage
\section{Contact group assignment}
\label{contact_groups}
To assign individuals to contact groups, we inferred likely groups from their UrbanPop attributes. PUMS household identifiers were included as an agent variable within the UrbanPop synthetic population. These identifiers were used to assign individuals to household groups.
\\\\
School groups were assigned to students based on their school identifier and school grade. School grades ranged from pre-kindegarten through to tertiary level. Target school group sizes for pre-kindegarten, kindegarten, elementary, and secondary school settings were derived from student-to-teacher ratios for each county in New Mexico \cite{alexander2025epicast}. We limited the maximum student-to-teacher ratio to 50. For tertiary settings, we assumed a target group size of 37.5 individuals based on a national university class size estimate \cite{uni_est}. Where the number of students attending a particular school and sharing the same school grade exceeded the target school group size by more than $1.5$ times the target size, we divided the students into smaller groups to more closely match the target school group size. For example, for 65 grade 1 students assigned to the same school which is located in a county with a mean elementary student-to-teacher ratio of 15, we would compute $65/15=4.33$, which would then be rounded to the nearest whole number (i.e., 4 school groups). We then randomly assigned students to these designated school groups. We assigned one individual who is employed in educational services (North American Industry Classification System (NAICS): 611), and has a daytime census blockgroup matching a given school, as a teacher in each school group.
\\\\
To assign work groups in the population, we grouped together employed individuals with the same industry of employment (designated by their NAICS code), and the same daytime census blockgroup. Similar to school groups, we embedded heterogeneity in group sizes through target work group sizes for each industry, derived from the 2019 County Business Patterns (CBP) estimates of establishment size by industry \cite{alexander2025epicast}.
\\\\
Finally, each individual was assigned a daytime and nighttime community group equivalent to the daytime and nighttime census blockgroups assigned to them in the UrbanPop synthetic population. Individuals within the same work groups and school groups shared the same daytime community group. Similarly, individuals in the same household shared the same nighttime community group. While all agents were assigned a nighttime census blockgroup within New Mexico, a small portion of the population was assigned a daytime census blockgroup outside of New Mexico, representing commuters travelling interstate. To ensure the agent population and associated contact groups was a closed system, we re-assigned these agents to daytime census blockgroups within New Mexico. Blockgroups for reassignment were chosen within the county encompassing the agent's nighttime census blockgroup.

\clearpage
\section{Contact sampling}
\label{contact_sampling}
The parameter values chosen for $c_k$ were derived from the POLYMOD study \cite{mossong2008social}, a set of empirical contact surveys conducted in Europe, measuring the number of unique contacts that individuals had in different transmission setting across a survey day. Transmission settings in POLYMOD that were not households, schools or workplaces (e.g., transport, leisure) were classified here as community contacts. We assumed community contacts were split evenly between daytime and nighttime settings. 
\\\\
The proportion of contact occurring in different transmission settings in POLYMOD provide an estimate of the relative contribution of each transmission setting to the overall contact rate (e.g., 23\% of contacts occur in households). We assumed all household groups were fully connected and household contact was limited only to individuals who shared a household (Equation 2). These two assumptions enforced a household contact rate of 2.247 contacts per day. To align with the relative contribution of each transmission setting to the overall contact rate in POLYMOD, we set the overall contact rate (i.e., $c=\sum_k c_k$) by dividing the household contact rate by the relative contribution of household contact (i.e., $c=2.247/0.23=9.77$ contacts per day). We then used the relative contribution of the other transmission settings to define the setting specific contact rate in the remaining settings.
\\\\
For each transmission setting, we sampled $0.5 \times c_k \times |A|$ (rounded to the nearest integer) node pairs without replacement. The $\frac{1}{2}$ is included in the product as each sampled edge increases the degree of two nodes by 1. Thus, to align the mean degree of nodes with the contact rate $c$, the number of sampled edges must be halved. We sampled without replacement to ensure contacts were unique, aligning with the definition typically used in contact surveys. Note that we sampled without replacement across transmission settings to avoid an edge being sampled twice in different transmission settings. To do this, after an edge $E(A_i,A_j)$ is sampled, $P_k(E(A_i,A_j))$ and $P_k(E(A_j,A_i))$ are set to zero in all transmission settings, and each $P_k$ is re-normalised. As individuals are able to share multiple contact groups with others (e.g., household members are in the same nighttime community group by definition), the order of transmission settings that are considered when sampling edges can be significant. We sampled edges from transmission settings in the following order based on the significance of the contact settings: households, workplaces, schools, nighttime communities, and daytime communities. 

\clearpage
\section{Age estimation bias}
\label{age_est}
\begin{figure}[h]
    \centering
    \includegraphics[width=0.75\linewidth]{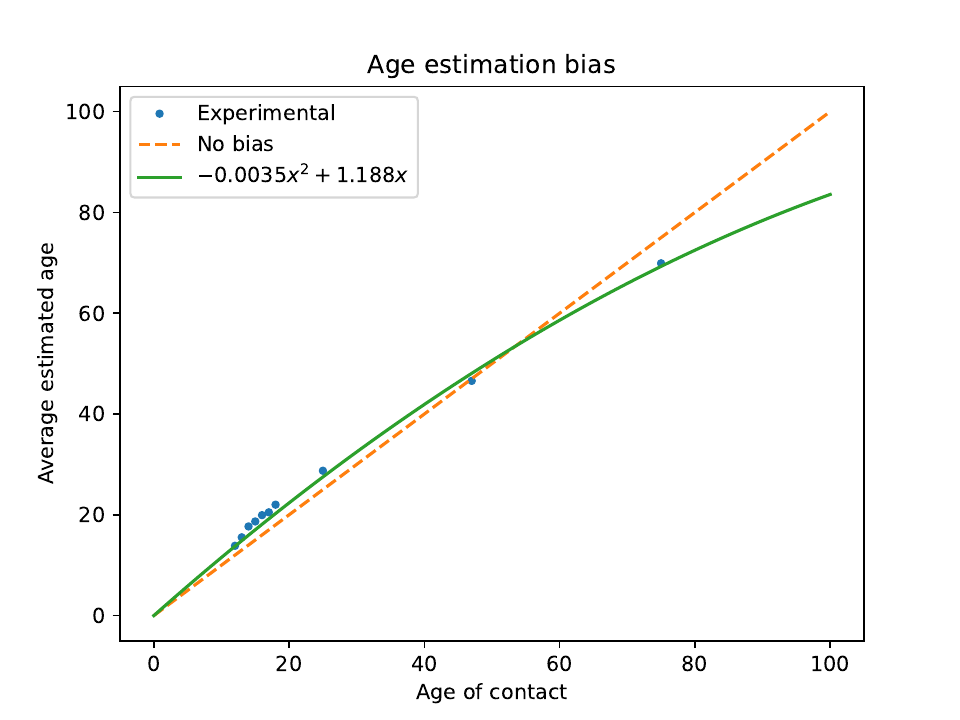}
    \caption{Age estimation bias. Experimental data (dots) collected from age estimation studies \cite{voelkle2012let,norja2022old} informs the estimation of age in simulated surveys. Fitting a quadratic polynomial to the experimental data (green) describes a general trend of overestimating the age of adolescents and young adults, and underestimating the age of older individuals. The fitted quadratic formula is used to inform the mean age estimate of survey participants recalling contact with any individual.}
    \label{fig:age_bias}
\end{figure}

To encode bias when estimating another individual's age, we defined a function $B(\text{age}_j)$ that defines the expected age estimate given a subject's age. When a contact survey participant estimates the age of a contact, they sampled an age from a Gamma distribution $P_{\text{age}}$. We used $B(\text{age}_j)$ in the definition of the mean of that gamma distribution (Equation 3).
\\\\
To define $B(\text{age}_j)$, we fit a quadratic polynomial to experimental estimates of age estimation bias (Figure \ref{fig:age_bias}). We relied on two studies of age estimation accuracy. The first asked participants of different ages and genders to estimate the age of subject faces based on photographs that varied by age, gender, and facial expression \cite{voelkle2012let}. The second study focused specifically on the estimation of younger (12-18 years old) individuals, asking participants of various ages and genders to estimate the age based on facial photos \cite{norja2022old}. We extracted average age estimates given a subject age from both studies (see blue markers below).
We use the scaling factor $a$ to modulate the effect of the age estimation bias (see Equation 3). Experimental conditions in studies of age estimation bias are most similar to the community transmission setting. When $a$ is set to $1/0.39=2.56$, we ensured the product of $\epsilon_Ca$ (i.e., the bias multiplier for community contacts) was equal to 1, and therefore $\mu=B(\text{age}_j)$. As such, we chose $a=2.56$ as a representative value our study that most closely aligns with the empirically observed bias. Similarly, we set $h$ (constant describing the maximum distribution variance) to 10 in our main analysis to align with experimental variation. Specifically, distribution variance in community contact settings (i.e. $\sigma^2=\epsilon_Ch=3.9$) produced variation similar to variation reported in age estimation standard deviation in \cite{norja2022old}. 
\\\\
We assumed the mean of $P_{\text{age}}$ should be monotonically increasing with contact age (i.e., the mean age estimate should never decrease for an older contact). We defined a range of $a$ where this was ensured for all ages $\leq110$ years. To do this, we took the derivative of the function $f(x)$ for deriving our mean age estimate $\mu$ (Equation 3):
\begin{equation*}
    B(x)=a'x^{2}+b'x
\end{equation*}
\begin{equation*}
    f(x)=x+\epsilon_ka(a'x^{2}+b'x-x)
\end{equation*}
\begin{equation*}
    f'(x)=1+\epsilon_ka(2a'x+b'-1)
\end{equation*}
To calculate the value for $a$ associated with the turning point of $f(x)$, we solved for when $f'(x)=0$:
\begin{equation*}
    0=1+\epsilon_ka(2a'x+b'-1)
\end{equation*}
\begin{equation*}
    a=\frac{-1}{\epsilon_k(2a'x+b'-1)}
\end{equation*}
Finally, assuming the fitted values for $a'$ and $b'$ from Figure \ref{fig:age_bias} above ($a'=-0.0035$, $b'=1.188$), and the highest value for $\epsilon_k$ when the bias is most active (which is found in the work context $\epsilon_w=0.44$), we solve for when the turning point is at $x=110$, which :
\begin{equation*}
    {a=3.87}
\end{equation*}
Therefore, $0\leq a\leq3.87$ to ensure a monotonically increasing age estimation for all contact ages $\leq110$ years.

\clearpage
\section{Impact of within-group racial bias among White survey participants}
\label{wg_bias}
\begin{figure}[h]
    \centering
    \includegraphics[width=\linewidth]{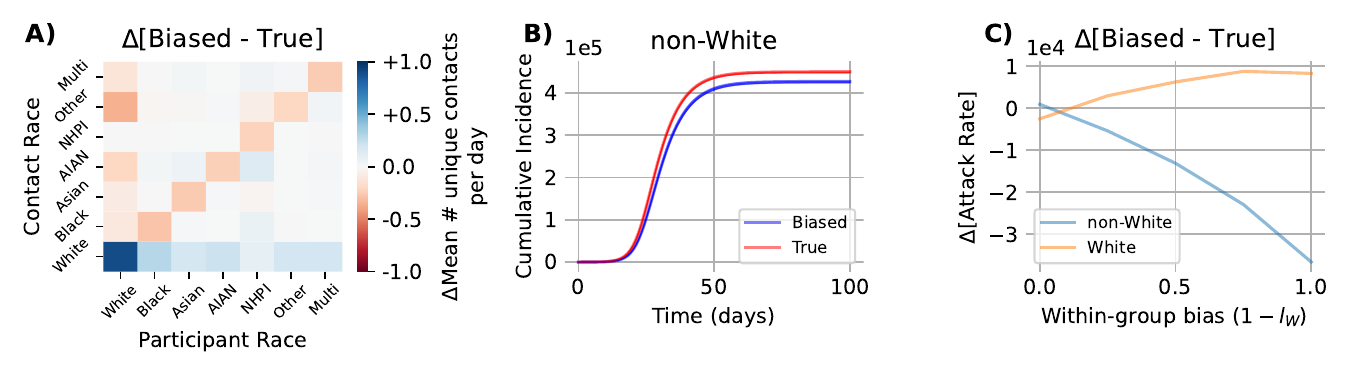}
    \caption{Impact of within-group bias affecting the estimation of contact race on estimated contact patterns (A) and epidemic dynamics (B \& C). Racial identification of others may be influenced by the race of the observer. By comparing the average contact matrix from ten simulated surveys with biased (`Biased') and unbiased (`True') recall, within-group bias in the racial identification of others among White survey participants ($l_{\text{W}}=0.25$) leads to systematic misidentification of non-White individuals as White (A). Using the biased contact matrix to simulate disease spread in a SIR model ($R_{0}=2.9$), underestimates the cumulative incidence among non-White individuals (B). Increasing the bias magnitude leads to an approximately linear decrease in attack rate among non-White individuals, and an increase in attack rate among White individuals (C).}
    \label{fig:fig5}
\end{figure}

To capture the potential effects of within-group bias on racial identification accuracy, we expanded the definition of $P_{\text{race}}$ given the contact’s race $\text{race}_j$, and the transmission setting $k$ and census tract $g$ in which the contact occurred as:
\begin{equation}
    P_{\text{race}}(x=b\,|\,\text{race}_i= \text{race}_j,k,g,r_b,l_a)=\begin{cases}
        1-(l_a-l_a\zeta_{g,b}) \epsilon_{k} r_b&,\,\,\text{if }b=\text{race}_i=\text{race}_j \\
        l_a\zeta_{g,b}\epsilon_{k} r_b &,\,\,\text{otherwise}
    \end{cases}
\end{equation}
\begin{equation}
    P_{\text{race}}(x=b\,|\,\text{race}_i\neq \text{race}_j,k,g,r_b,l_a)=\begin{cases}
        1-(1-l_a\zeta_{g,b}) \epsilon_{k} r_b&,\,\,\text{if }b=\text{race}_j \\
        (1-l_a+ l_a\zeta_{g,b})\epsilon_{k} r_b&,\,\,\text{if }b= \text{race}_i \\
        l_a\zeta_{g,b}\epsilon_{k} r_b&,\,\,\text{otherwise}
    \end{cases}
\end{equation}
where $b \in$ \{White, Black, Asian, AIAN, Native Hawaiian/Pacific Islander, Other, Multiracial\}, $r_b$ is a scaling factor for recall accuracy associated with contacts identifying with race $b$, $\epsilon_{k}$ is the transmission setting scaling factor for setting $k$ in which the interaction occurred, $\zeta_{g,b}$ is the proportion of the population identifying as race $b$ in the census tract $g$ where the interaction occurred, and $l_a$ represents the extent to which within-group bias experienced by the participant ($A_i$) who identifies with race $a$ (i.e., $\text{race}_i=a$) affects estimation.
\\\\
We simulated a within-group bias affecting White participants by varying $l$ for White participants ($l_{\text{W}}$ for all $\text{race}_i=\text{White}$) between 0 and 1, where 0 represents maximum within-group bias and 1 represents no within-group bias. We set the participant bias for all non-White participants ($l_{\text{NW}}$ for all $\text{race}_i\neq\text{White}$) to 1 (i.e., no within-group bias), and set the recall accuracy scaling factor $r_b$ for all contacts (regardless of race) to 2.05.
\\\\
We found embedding a within-group racial bias among White survey participants ($l_{\text{W}}=0.25$) lead to a reduction in reported contact frequency with non-White individuals (Figure \ref{fig:fig5}A). When comparing the average contact matrix from ten random samples ($\text{Sample Size}=10,000$) of the population, where we assume separately biased (`Biased') and unbiased (`True') contact race recall, White participants reported on average 0.9 fewer contacts with individuals identifying as non-White, misidentifying these contacts as White. Non-White contact assortativity decreased reflecting the impact of census tract racial demographics informing race estimation. Reduction in the reported non-White contact rates lead to a decrease in cumulative incidence among non-White individuals using an SIR model (Figure \ref{fig:fig5}B). As the magnitude of within-group estimation bias ($1-l_{\text{W}}$) increased, the attack rate among the non-White individuals decreased, and among White individuals increased (Figure \ref{fig:fig5}C), reflecting the changes in contact rates.

\clearpage
\section{Visualisation of contact estimation process}
\label{eg_est}
\begin{figure}[h]
    \centering
    \includegraphics[width=1\linewidth]{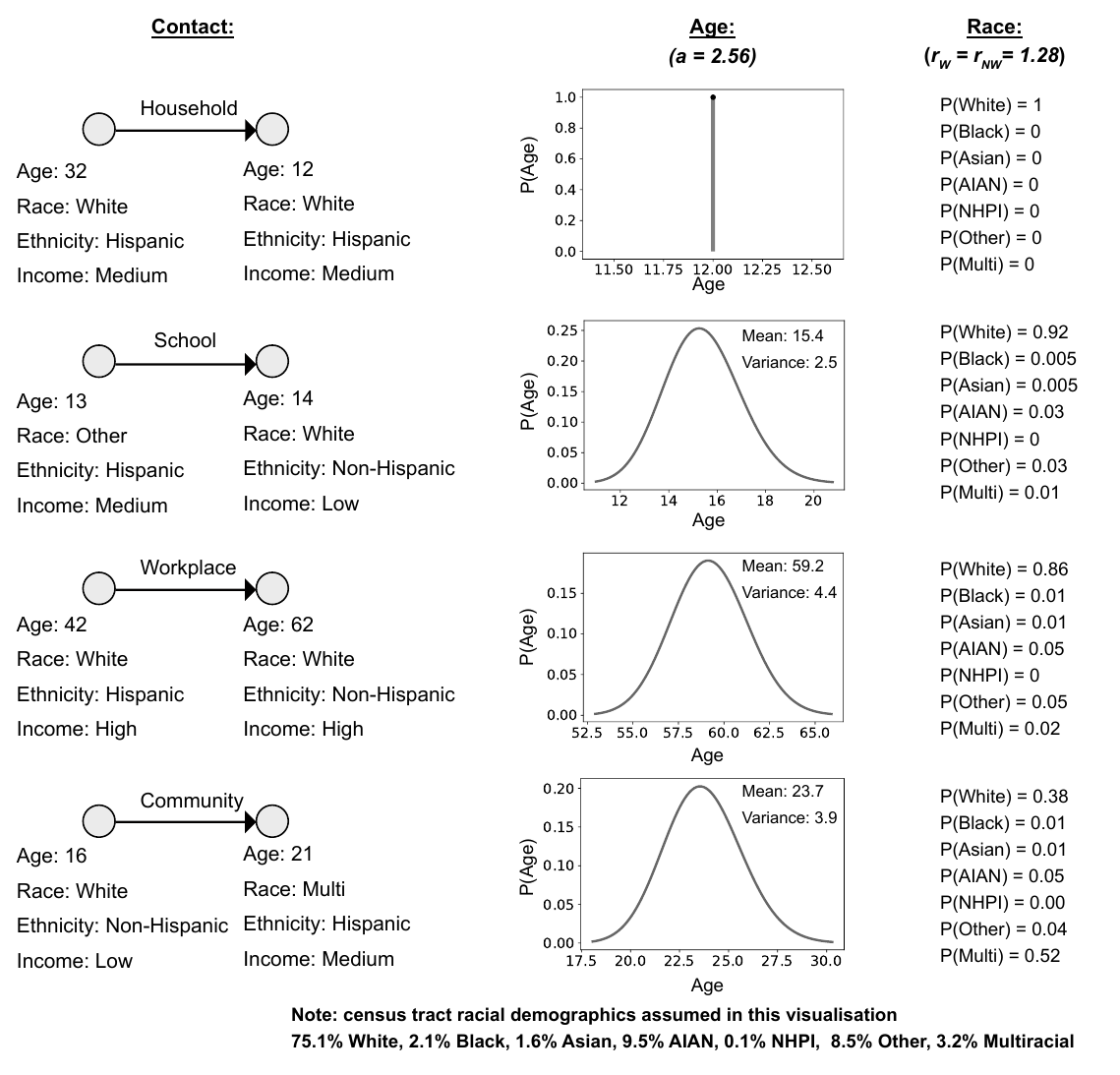}
    \caption{Visualisation of contact age and race estimation process in survey simulation. For each contact, a survey participant will sample an estimate of age from a continuous probability distribution (`Age' column) and race from a categorical probability distribution (`Race' columns). The mean and variance of the age distribution will depend on the true age of the contact ($\text{age}_j$), the setting of the contact ($\epsilon_k$) and the magnitude of bias ($a$) when estimating contacts of this age (Equations 3 \& 4). The likelihood of estimating a particular race will depend on the true race of the contact ($\text{race}_j$), the setting of the contact ($\epsilon_k$), and the assumed extent of racial perception bias for race $b$ ($r_b$; Equation 5).}
    \label{fig:estimation_vis}
\end{figure}

\clearpage
\section{Estimating contact ethnicity ($P_{\text{ethnicity}}$)}
\label{eth_est}
\begin{figure}[h]
    \centering
    \includegraphics[width=\linewidth]{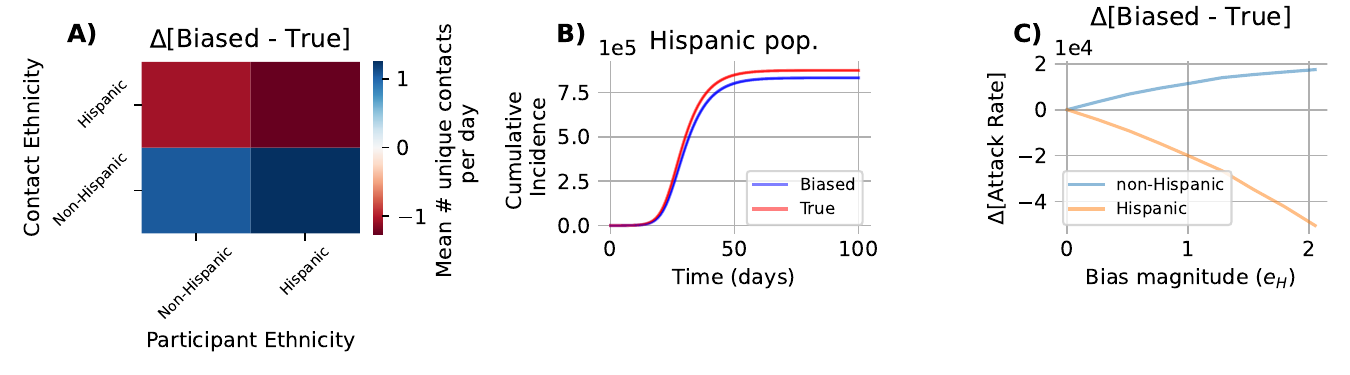}
    \caption{Impact of bias affecting the estimation of contact ethnicity on estimated contact patterns (A) and epidemic dynamics (B \& C). Ethnic misidentification typically occurs at a higher rate for Hispanic individuals. By comparing the average contact matrix from ten simulated surveys with biased (`Biased') and unbiased (`True') recall, ethnic bias in the identification of Hispanic individuals ($e_{\text{H}}=1.54$) leads to systematic misidentification of Hispanic individuals as non-Hispanic (A). Using the biased contact matrix to simulate disease spread in a SIR model ($R_{0}=2.9$), underestimates the cumulative incidence among Hispanic individuals (B). Increasing the bias magnitude leads to an approximately linear decrease in attack rate among Hispanic individuals, and an increase in attack rate among non-Hispanic individuals (C).}
    \label{fig:supp_eth}
\end{figure}

The accuracy of observer-identification when estimating ethnicity has been observed to vary depending on the ethnicity of the subject \cite{gomez2006misclassification}. Misidentification of Hispanic individuals' ethnicity typically occurs at a higher rate than non-Hispanic individuals, occurring in healthcare settings at rates as high as 59\% \cite{johnson2023accuracy}. To capture this difference between ethnic groups, we defined $P_{\text{ethnicity}}$, given the contact’s ethnicity ($\text{ethnicity}_j$), and the transmission setting $k$ and census tract $g$ in which the contact occurred as below:
\begin{equation}
    P_{\text{ethnicity}}(x=b\,|\, \text{ethnicity}_j,k,g,e_b)=\begin{cases}
        1-\epsilon_{k} e_b+\zeta_{g,b}\epsilon_{k} e_b &,\,\text{if }b=\text{ethnicity}_j \\
        \zeta_{g,b}\epsilon_{k} e_b &,\, \text{otherwise}
    \end{cases}\,,
\end{equation}
where $b \in$ \{Hispanic, non-Hispanic\}, $e_b$ is a scaling factor for recall accuracy associated with contacts identifying with ethnicity $b$, and $\zeta_{g,b}$ is the proportion of the population identifying as ethnicity $b$ in census tract $g$ where the interaction occurred. For $e_{\text{H}}$, the recall accuracy associated with Hispanic contacts, we chose values between 0 and 2.05. For $e_{\text{NH}}$, the recall accuracy associated with non-Hispanic contacts, we set to 0. These parameter settings replicated the conditions used in the main analysis of non-White estimation bias.
\\\\
We found bias affecting the estimation of contact ethnicity ($e_{\text{H}}=1.54$) lead to a reduction in reported contact frequency with Hispanic individuals (Figure \ref{fig:supp_eth}A). When comparing the average contact matrix from ten random samples ($N_{\text{sample}}=10,000$) of the population, where we assume separately biased (`Biased') and unbiased (`True') contact ethnicity recall, all participants typically reported between 0.7 and 0.9 fewer contacts with Hispanic individuals. Reduction in the reported Hispanic contact rates lead to a decrease in cumulative incidence among Hispanic individuals using an age- and ethnicity-stratified SIR model (Figure \ref{fig:supp_eth}B). As the magnitude of ethnicity estimation bias ($e_{\text{H}}$) increased, the attack rate among the Hispanic individuals decreased, and among non-Hispanic individuals increased (Figure \ref{fig:supp_eth}C), reflecting the changes in contact rates.

\clearpage
\section{Estimating contact income ($P_{\text{income}}$)}
\label{income_est}
\begin{figure}[h]
    \centering
    \includegraphics[width=\linewidth]{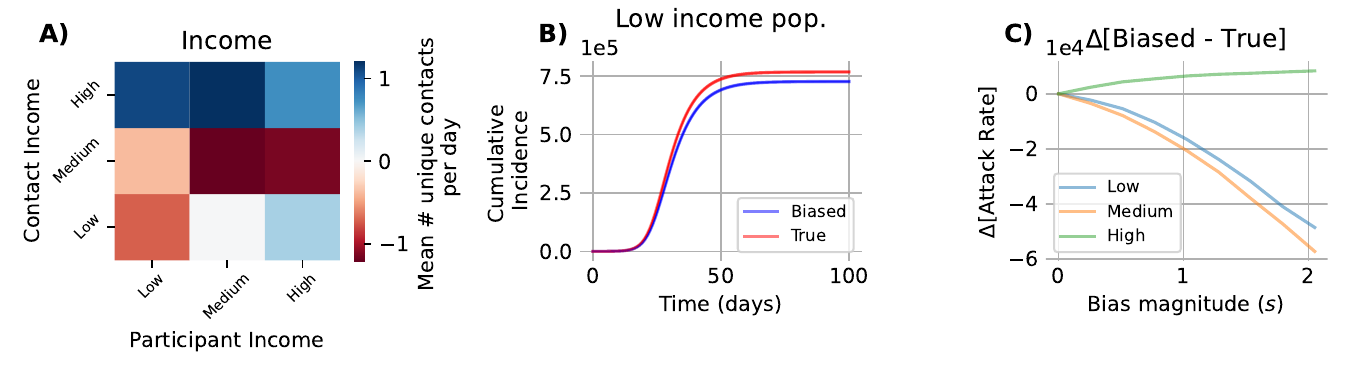}
    \caption{Impact of bias affecting the estimation of contact income stratum on estimated contact patterns (A) and epidemic dynamics (B \& C). Socio-economic status (SES) misidentification, a related individual measure to income, occurs at a substantial rate when interpreting visual and audio cues of strangers. By comparing the average contact matrix from ten simulated surveys with biased (`Biased') and unbiased (`True') recall, a general bias in the identification of contact income ($r_{\text{I}}=1.54$) leads to systematic misidentification of Low and Medium income individuals as High income (A). Using the biased contact matrix to simulate disease spread in a SIR model ($R_{0}=2.9$), underestimates the cumulative incidence among Low income individuals (B). Increasing the bias magnitude leads to a decrease in attack rate among Low and Medium income individuals, and a slight increase in attack rate among High income individuals (C).}
    \label{fig:supp_income}
\end{figure}

Similar to race and ethnicity, accurately estimating another individual's socio-economic status (SES) based on visual and audio cues has been shown to be difficult in empirical studies \cite{kraus2017signs,kraus2009signs}. Studies of observer-identification of SES have found the accuracy of determining SES in strangers is marginally better than making a random choice \cite{kraus2017signs}. For example, observers were found to identify the correct SES quartile in subjects around 35\% of the time across four different studies of SES identification accuracy. This inaccuracy was observed for low, medium and high SES individuals. The factors affecting a person's ability to estimate another individuals SES are not well understood. We used these empirical findings of SES identification to inform the estimation process of a contact's income stratum, a related individual measure.
\\\\
To capture inaccuracy in a survey participant's estimation of a contact's income stratum, we defined $P_{\text{income}}$ given the contact’s income ($\text{income}_j$), and the transmission setting $k$ in which the contact occurred as below:
\begin{equation}
    P_{\text{income}}(x=b\,|\, \text{income}_j,k,s)=\begin{cases}
        1-\epsilon_{k} s+\frac{1}{3}\epsilon_{k} s &,\,\text{if }b=\text{income}_j \\
        \frac{1}{3}\epsilon_{k} s &,\, \text{otherwise}
    \end{cases}\,,
\end{equation}
where $b \in$ \{Low, Medium, High\}, and $s$ is a scaling factor for recall accuracy associated with identifying any contact's income stratum $b$. For $s$, we chose values between 0 and 2.05. Instead of the geographic (census tract or state) bias used to inform estimation of race and ethnicity ($\zeta_{g,b}$), we used the ratio $1/3$ to capture an evenly distributed bias between the three income strata. This definition was chosen based on there being no evidence that any particular SES category was misidentified at a higher rate than any other, or that estimation accuracy varied by geographic region.
\\\\
We found bias affecting the estimation of contact income ($s=1.54$) lead to a reduction in reported contact frequency with Low and Middle income individuals (Figure \ref{fig:supp_income}A). When comparing the average contact matrix from ten random samples ($N_{\text{sample}}=10,000$) of the population, where we assume separately biased (`Biased') and unbiased (`True') contact income recall, participants typically reported fewer contacts with Low and Middle income individuals, and more contact with High income individuals. Reduction in the reported Low income contact rates lead to a decrease in cumulative incidence among Low income individuals using an age- and income-stratified SIR model (Figure \ref{fig:supp_income}B). As the magnitude of income estimation bias ($s$) increased, the attack rate among the Low and Middle income individuals decreased, and among High income individuals increased marginally (Figure \ref{fig:supp_eth}C), reflecting the changes in contact rates. 

\clearpage
\section{Additional SIR model details}
\label{ngm}
We defined the age-specific susceptibility for group \textbf{a} ($\beta_{\textbf{a}}$) as:
\begin{equation}
    \beta_{\textbf{a}} = \kappa\beta'_{\textbf{a}}\,,
\end{equation}
where $\kappa$ is some transmission scaling constant, and $\beta'_{\textbf{a}}$ is the relative age-specific susceptibility for individuals in group $\textbf{a}$ to infection. 
\\\\
We set the value of $\kappa$ such that the basic reproduction number ($R_0$) (the average number of secondary infections produced by an infected individual in a completely susceptible population) of the SIR model was equivalent to a target $R_0$ value. We computed the $R_0$ of the SIR model as the dominant eigenvalue of the Next Generation Matrix (NGM) (section \ref{ngm_extra}).
\\\\
For each simulation, we seeded an outbreak by assuming 0.01\% of the population was infected, proportionally distributed across the population strata. We set $\kappa$ (transmission scaling constant) to achieve a specific $R_0$ varying between 1.2 and 6, with particular focus on $R_0=2.9$ (similar to COVID-19 \cite{billah2020reproductive}). We set $\gamma$ (recovery rate), and $\beta'$ (age-specific relative susceptibility) based on values estimated for COVID-19 spread (Table \ref{tab:sir_params_main}) \cite{davies2020age}.
\begin{table}[h]
\centering
\resizebox{\columnwidth}{!}{%
\begin{tabular}{|l|l|l|l|l|l|l|l|l|l|l|l|}
\hline
\textbf{Pathogen} & \textbf{$\boldsymbol{\gamma}$} & \textbf{$\boldsymbol{\beta'_{[0,5]}}$} & \textbf{$\boldsymbol{\beta'_{[6,10]}}$} & \textbf{$\boldsymbol{\beta'_{[11,20]}}$} & \textbf{$\boldsymbol{\beta'_{[21,30]}}$} & \textbf{$\boldsymbol{\beta'_{[31,40]}}$} & \textbf{$\boldsymbol{\beta'_{[41,50]}}$} & \textbf{$\boldsymbol{\beta'_{[51,60]}}$} & \textbf{$\boldsymbol{\beta'_{[61,65]}}$} & \textbf{$\boldsymbol{\beta'_{[66,70]}}$} & \textbf{$\boldsymbol{\beta'_{>70}}$} \\ \hline
SARS-CoV-2 & 0.2 & 0.4 & 0.4 & 0.38 & 0.79 & 0.86 & 0.8 & 0.82 & 0.88 & 0.88 & 0.74 \\ \hline
\end{tabular}
}
\caption{Parameter settings for age-structured SIR model \cite{davies2020age}.}
\label{tab:sir_params_main}
\end{table}
\subsection{Deriving $R_0$ from the Next Generation Matrix (NGM) of a stratified SIR model}
\label{ngm_extra}
NGMs relate the number of newly infected individuals in various compartments across consecutive generations \cite{diekmann2010construction}. NGMs can be used to derive the basic reproductive number for an infectious disease system. Specifically, we can determine $R_0$ by computing the dominant eigenvalue of an NGM. 
\\\\
Consider the structured SIR model described in Equations 7, 8 and 9. When only considering two groups in a population, we first derive the linearised infection subsystem, describing just the production of new infections and changes in the states of individuals already infected. We assume some constant ratio $p$ for the total number of individuals in group 1 compared to 2 (i.e., $p=\frac{N_1}{N_2}$). We consider the above system at the infection-free steady state, where $I=R=0$ and $S_{i}=N_{i}$, reducing the above system of ODEs to:
\begin{equation}
    \frac{dI_1}{dt} = \beta_{1}G'_{11}I_{1} + p\beta_{1}G'_{12}I_{2}- \gamma_{1}I_{1},
\end{equation}
\begin{equation}
    \frac{dI_2}{dt} = \frac{1}{p}\beta_{2}G'_{21}I_{1} + \beta_{2}G'_{22}I_{2}- \gamma_{2}I_{2},
\end{equation}
where $I_\textbf{a}(t)$ is the population of infected individuals in group $\textbf{a}$ at time $t$; $G'_{\textbf{a,b}}$ is the per-capita contact rate between an individual in group $\textbf{a}$ with individuals in each other population group $\textbf{b}$; $\beta_{\textbf{a}}$ refers to the age-specific susceptibility of individuals in group $\textbf{a}$ to infection; $\gamma_{\textbf{a}}$ refers to the recovery rate specific to group $\textbf{a}$. Next, we derive the transmission matrix \textbf{T}, which describes the production of new infections (i.e., susceptible to infected) in the infected compartments:
\begin{equation}
\textbf{T}=
\begin{pmatrix}
\beta_{1}G_{11} & p\beta_{1}G_{12} \\\\
\frac{1}{p}\beta_{2}G_{21} & \beta_{2}G_{22} 
\end{pmatrix}
\end{equation}
We then derive the transition matrix \textbf{$\Sigma$}, which describes the changes of state in the infected compartments:
\begin{equation}
\textbf{$\Sigma$}=
\begin{pmatrix}
-\gamma_1 & 0 \\\\
0 & -\gamma_2 
\end{pmatrix}
\end{equation}
As \textbf{T} has no rows consisting entirely of zeros (see `Auxiliary matrix' in \cite{diekmann2010construction} for when this is not the case), we now only need to compute $-\textbf{T}\Sigma^{-1}$ to find the NGM $\textbf{K}$:
\begin{equation}
    \textbf{K}=-\textbf{T}\Sigma^{-1}=
    \begin{pmatrix}
    \frac{\beta_{1}G_{11}}{\gamma_1} &  \frac{p\beta_{1}G_{12}}{\gamma_2} \\\\
     \frac{\beta_{2}G_{21}}{p\gamma_1} &  \frac{\beta_{2}G_{22}}{\gamma_2} 
    \end{pmatrix}
\end{equation}
We can then find the $R_0$ of the above system by computing the dominant eigenvalue of $\textbf{K}$. 
\\\\
This can be generalised to any number of groups, where we define a matrix of population ratios ($p$) between groups by:
\begin{equation}
    p_{ij}=\frac{N_i}{N_j},
\end{equation}
where $N_i$ and $N_j$ are the total populations of individuals in groups $i$ and $j$, respectively.
\\\\
The NGM matrix (\textbf{K}) then becomes:
\begin{equation}
    \textbf{K}_{ij} = \frac{p_{ij} \beta_{i} G_{ij}}{\gamma_{i}}.
\end{equation}

\clearpage
\section{Setting-specific \& per-capita contact matrices}
\label{aux_matrices}
\begin{figure}[h]
    \centering
    \includegraphics[width=0.7\linewidth]{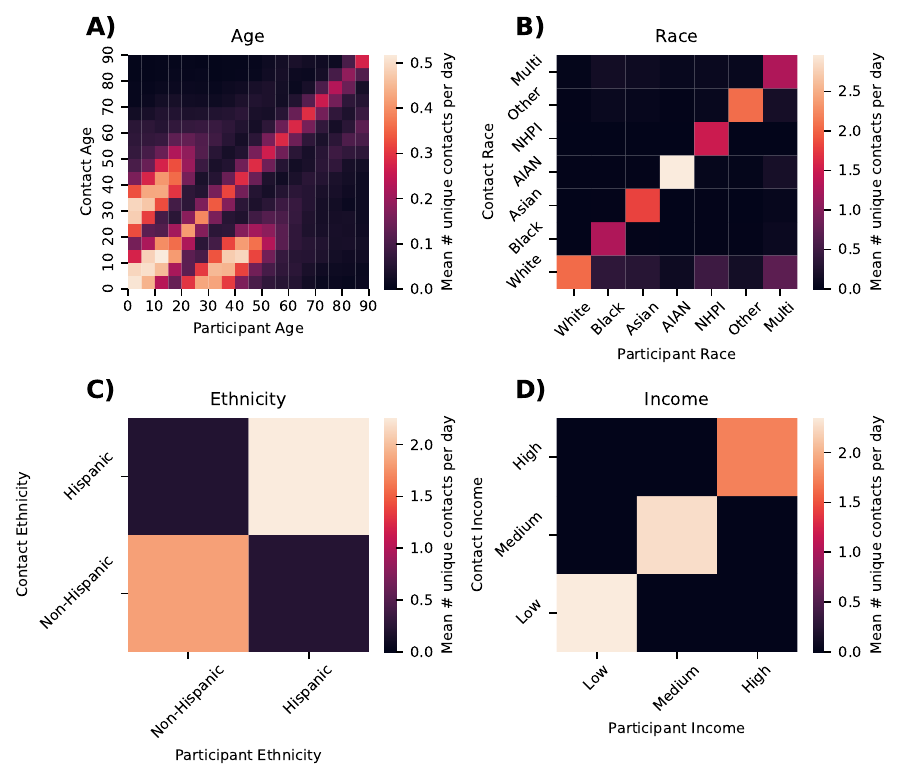}
    \caption{Household ground truth contact behaviour.}
    \label{fig:household}
\end{figure}

\begin{figure}[h]
    \centering
    \includegraphics[width=0.7\linewidth]{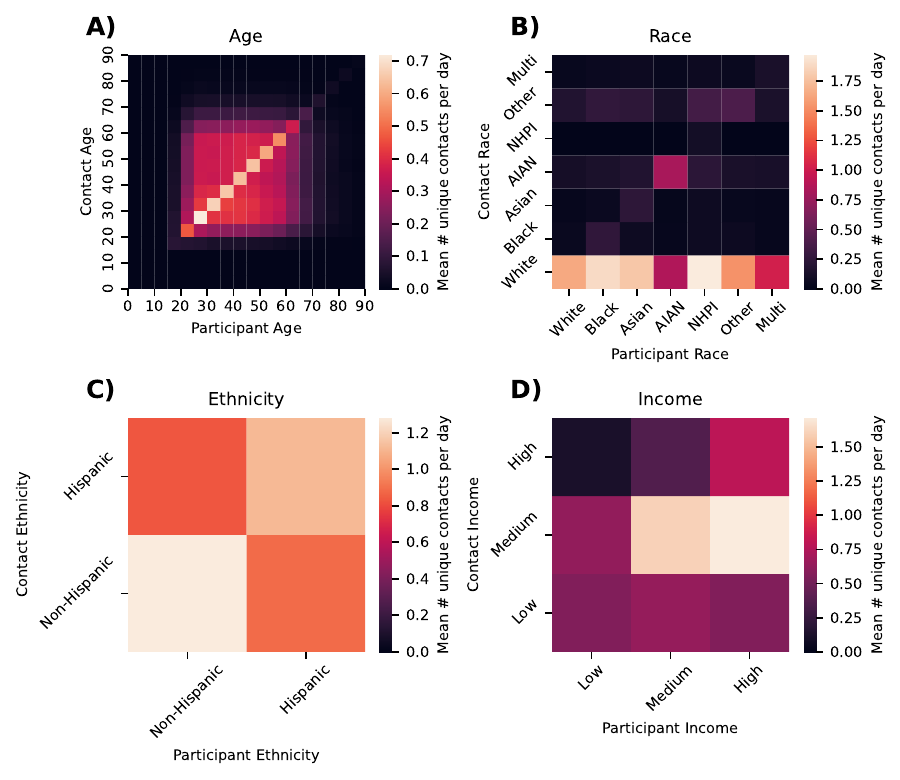}
    \caption{Workplace ground truth contact behaviour.}
    \label{fig:workplace}
\end{figure}

\begin{figure}[h]
    \centering
    \includegraphics[width=0.7\linewidth]{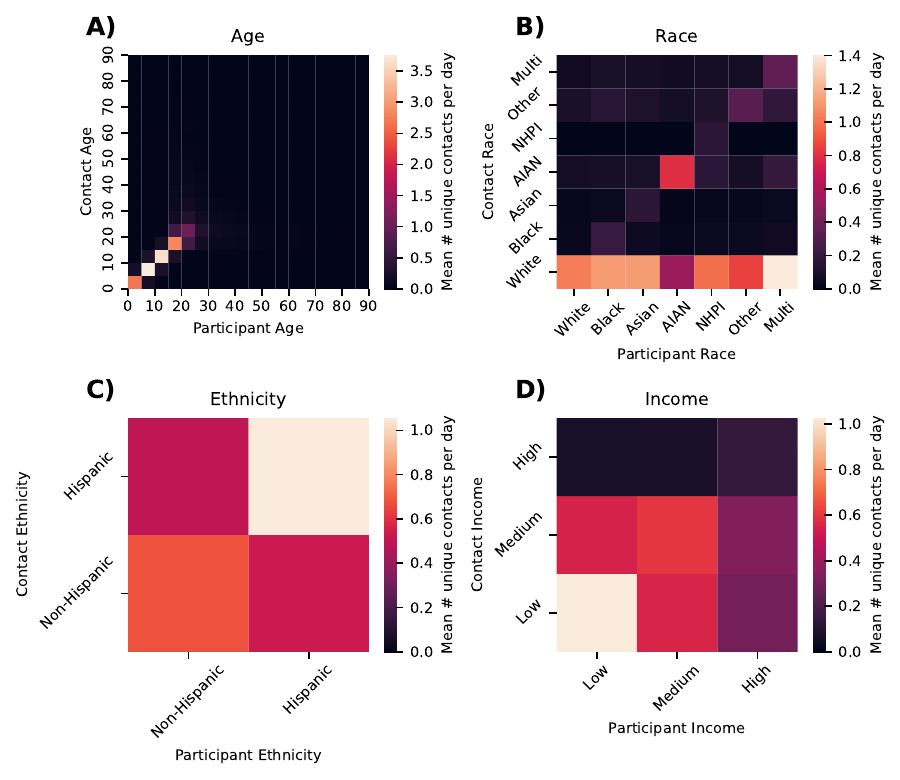}
    \caption{School ground truth contact behaviour.}
    \label{fig:school}
\end{figure}

\begin{figure}[h]
    \centering
    \includegraphics[width=0.7\linewidth]{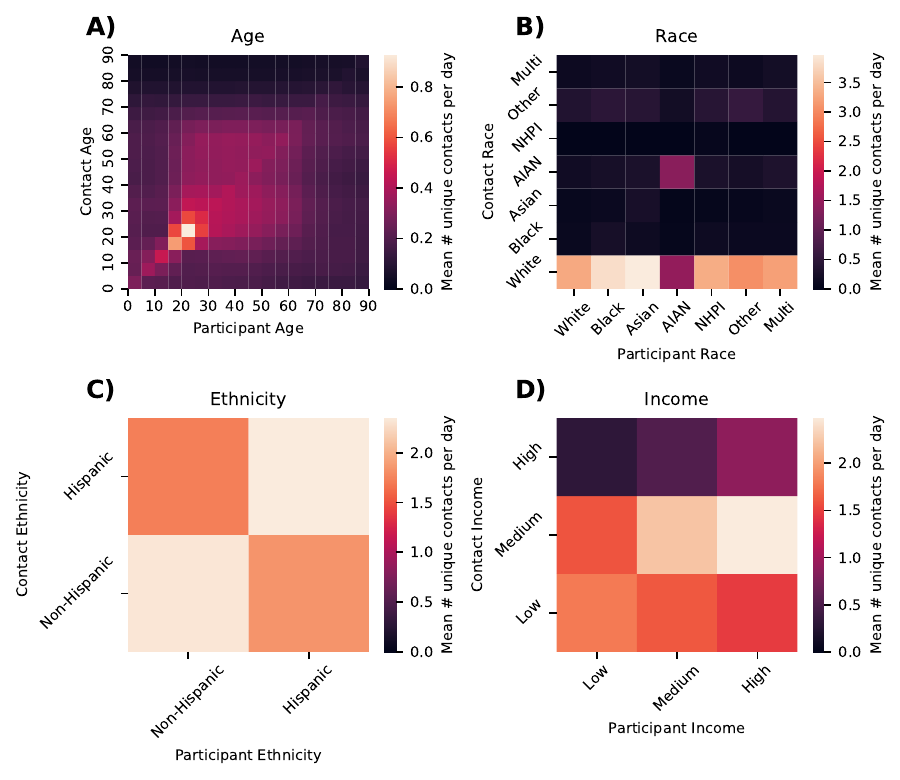}
    \caption{Community ground truth contact behaviour.}
    \label{fig:community}
\end{figure}

\begin{figure}[h]
    \centering
    \includegraphics[width=0.7\linewidth]{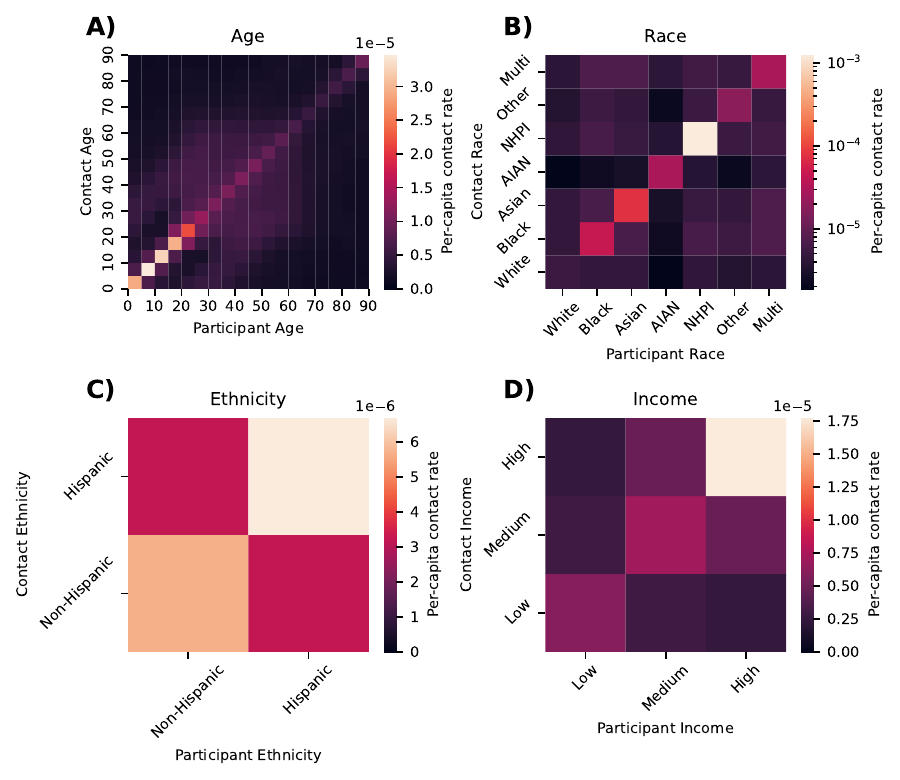}
    \caption{Per capita ground truth contact behaviour.}
    \label{fig:per_capita}
\end{figure}

To better understand how contact in the different transmission settings combine to form the overall contact behaviour described in Figure 2, we computed stratified contact matrices separately for contacts occurring in households (Figure \ref{fig:household}), workplaces (Figure \ref{fig:workplace}), schools (Figure \ref{fig:school}) and community settings (Figure \ref{fig:community}). Contact behaviour is generally assortative, ranging from weakly assortative in communities to strongly assortative in households. To adjust for the effects of the New Mexico population distribution, we computed the overall (i.e., all transmission settings) per-capita contact matrix adjusted for contact symmetry (Figure \ref{fig:per_capita}).

\clearpage
\section{Alternative visualisation of impact of perception biases on epidemic dynamics: disease prevalence}
\label{prev}
\begin{figure}[h]
    \centering
    \includegraphics[width=0.6\linewidth]{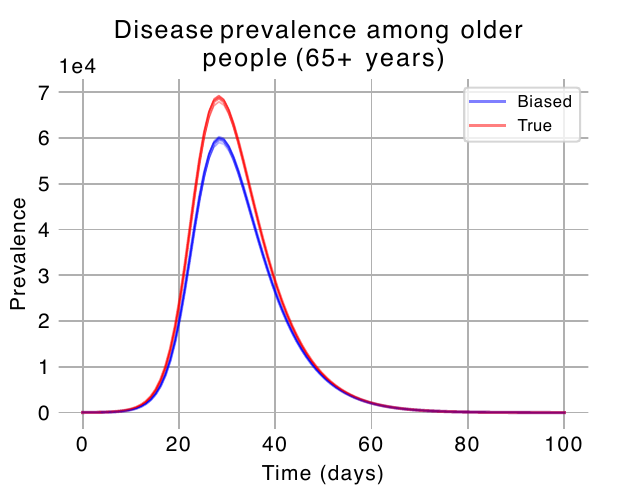}
    \caption{Impact of bias affecting the estimation of contact age ($a=2.56$) on epidemic dynamics. The age of older individuals is often underestimated when presented with limited visual and audio stimulus, similar to information available in casual contact settings. By comparing the average contact matrix from ten simulated surveys with biased (`Biased') and unbiased (`True') recall, surveys affected by age perception bias lead to an underestimation of reported contact with older individuals. Using an SIR model of disease spread ($R_{0}=2.9$), simulated epidemic dynamics which relied on biased survey data underestimated the true disease burden among people aged 65 years or older.}
    \label{fig:peak_prev}
\end{figure}

\begin{figure}[h]
    \centering
    \includegraphics[width=0.6\linewidth]{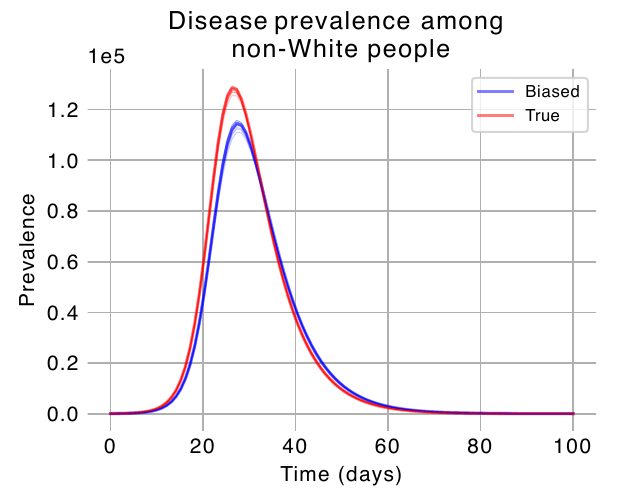}
    \caption{Impact of bias affecting the estimation of contact race ($r_\text{NW}=1.79$) on epidemic dynamics. Racial misidentification typically occurs at a higher rate for non-White individuals when observers are presented with limited visual and audio stimulus, similar to information available in casual contact settings. By comparing the average contact matrix from ten simulated surveys with biased (`Biased') and unbiased (`True') recall, surveys affected by race perception bias lead to an underestimation of reported contact with non-White individuals individuals. Using an SIR model of disease spread ($R_{0}=2.9$), simulated epidemic dynamics which relied on biased survey data underestimated the true disease burden among non-White individuals.}
    \label{fig:peak_prev_race}
\end{figure}

\clearpage
\section{Epidemic dynamics differences under $R_0=1.4$}
\label{lower_r0}
\begin{figure}[h]
    \centering
    \includegraphics[width=0.9\linewidth]{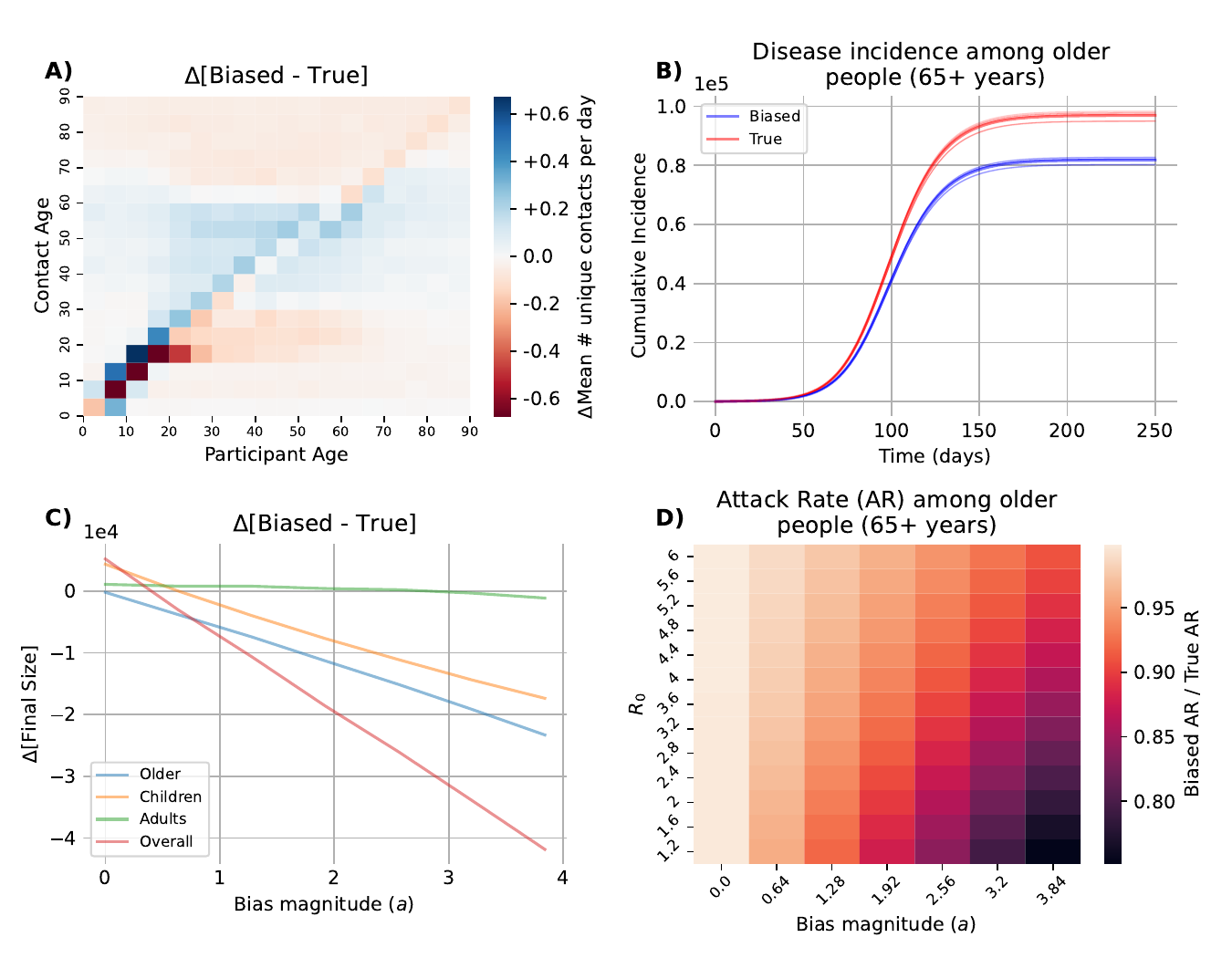}
    \caption{Impact of bias affecting the estimation of contact age on estimated contact patterns (A) and epidemic dynamics (B, C \& D). The age of older people (65+ years) is often underestimated by individuals when presented with limited visual and audio stimulus, similar to information available in casual contact settings. By comparing the average contact matrix from ten simulated contact surveys with biased (`Biased') and unbiased (`True') participant recall, surveys affected by age perception bias ($a=2.56$) lead to an underestimation of reported contact with older individuals (A). Using an SIR model of disease spread ($R_{0}=1.4$), simulated epidemic dynamics which relied on biased survey data underestimated the true disease burden among people aged 65 years or older (B). Increasing the magnitude of the age estimation bias ($a$) lead to greater underestimation of outbreak final size in children (0-19 years) and older people (65+ years) and a consistent estimation of outbreak final size in the adult population (20-64 years) (C). Differences in subpopulation final size when $a=0$ came about due to simulated variance in the age estimation process even when assuming no directional bias (Equations 3 \& 4). The ratio of the simulated attack rate (AR) using the biased contact survey and the true contact survey data (`Biased AR / True AR') decreased when assuming larger $a$ and a lower basic reproduction number (D).}
    \label{fig:fig3_low}
\end{figure}

\begin{figure}[h]
    \centering
    \includegraphics[width=\linewidth]{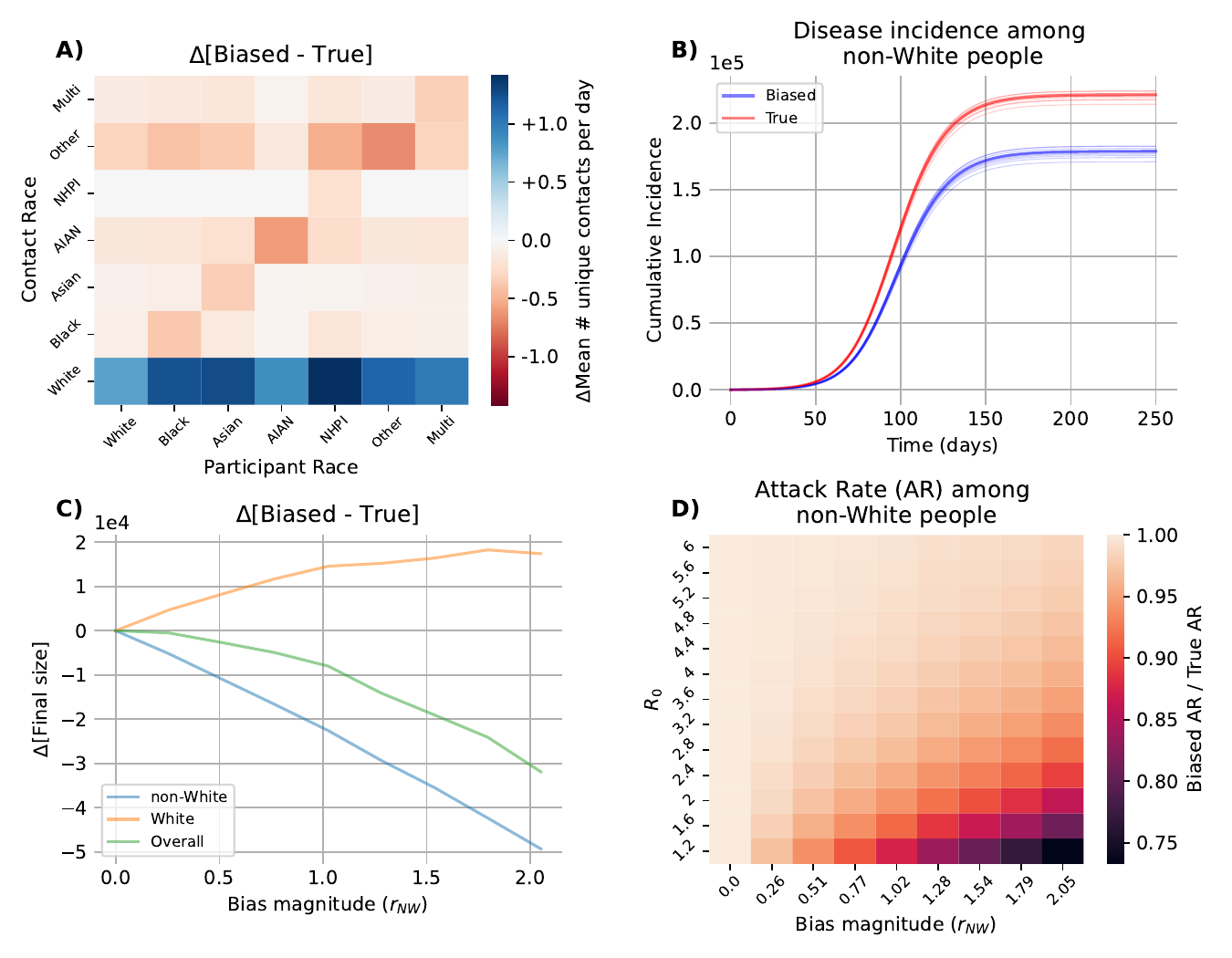}
    \caption{Impact of bias affecting the estimation of contact race on estimated contact patterns (A) and epidemic dynamics (B, C \& D). Racial misidentification typically occurs at a higher rate for non-White individuals when observers are presented with limited visual and audio stimulus, similar to information available in casual contact settings. By comparing the average contact matrix from ten simulated surveys with biased (`Biased') and unbiased (`True') recall, racial bias in the identification of non-White individuals ($r_{\text{NW}}=1.79$) leads to systematic misidentification of non-White individuals as White (A). Using the biased contact matrix to simulate disease spread in a SIR model ($R_{0}=1.4$), underestimates the cumulative incidence among non-White individuals (B). Increasing the bias magnitude leads to an approximately linear decrease in final size among non-White individuals, and an increase in final size among White individuals (C). The ratio of the simulated attack rate (AR) using the biased contact survey and the true contact survey data (`Biased AR / True AR') decreased when assuming larger $r_{\text{NW}}$ and a lower basic reproduction number (D).}
    \label{fig:fig4_low}
\end{figure}

\clearpage
\section{Epidemic dynamics differences under no age-specific susceptibility parametrisation}
\label{no_age}
\begin{figure}[h]
    \centering
    \includegraphics[width=1\linewidth]{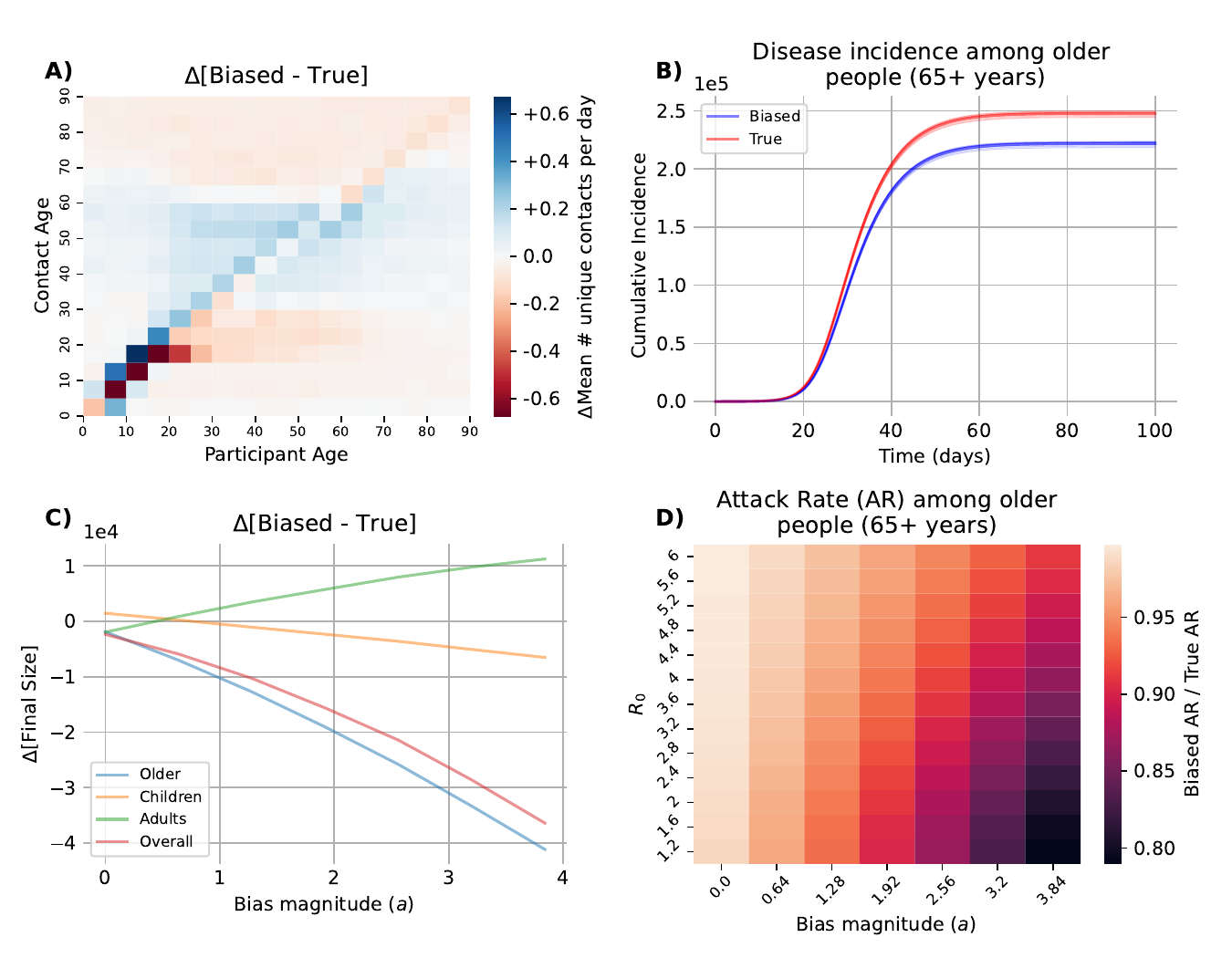}
    \caption{Impact of bias affecting the estimation of contact age ($a=2.56$) on estimated contact patterns (A) and epidemic dynamics (B, C \& D), parametrised with uniform age-specific susceptibility. The age of older individuals is often underestimated when observers are presented with limited visual and audio stimulus, similar to information available in casual contact settings. By comparing the average contact matrix from ten simulated surveys with biased (`Biased') and unbiased (`True') recall, surveys affected by age perception bias lead to an underestimation of reported contact with older individuals (A). Using an SIR model of disease spread ($R_{0}=2.9$), simulated epidemic dynamics which relied on biased survey data underestimated the true disease burden among people aged 65 years or older (B). Increasing the magnitude of age estimation bias ($a$) lead to an increasing underestimation of attack rate among children and older people and a consistent attack rate in the adult population (C). The ratio of the simulated attack rate (AR) using the biased contact survey and the true contact survey data (`Biased AR / True AR') decreased when assuming larger $a$ and a lower basic reproduction number (D).}
    \label{fig:no_age}
\end{figure}

\begin{figure}[h]
    \centering
    \includegraphics[width=1\linewidth]{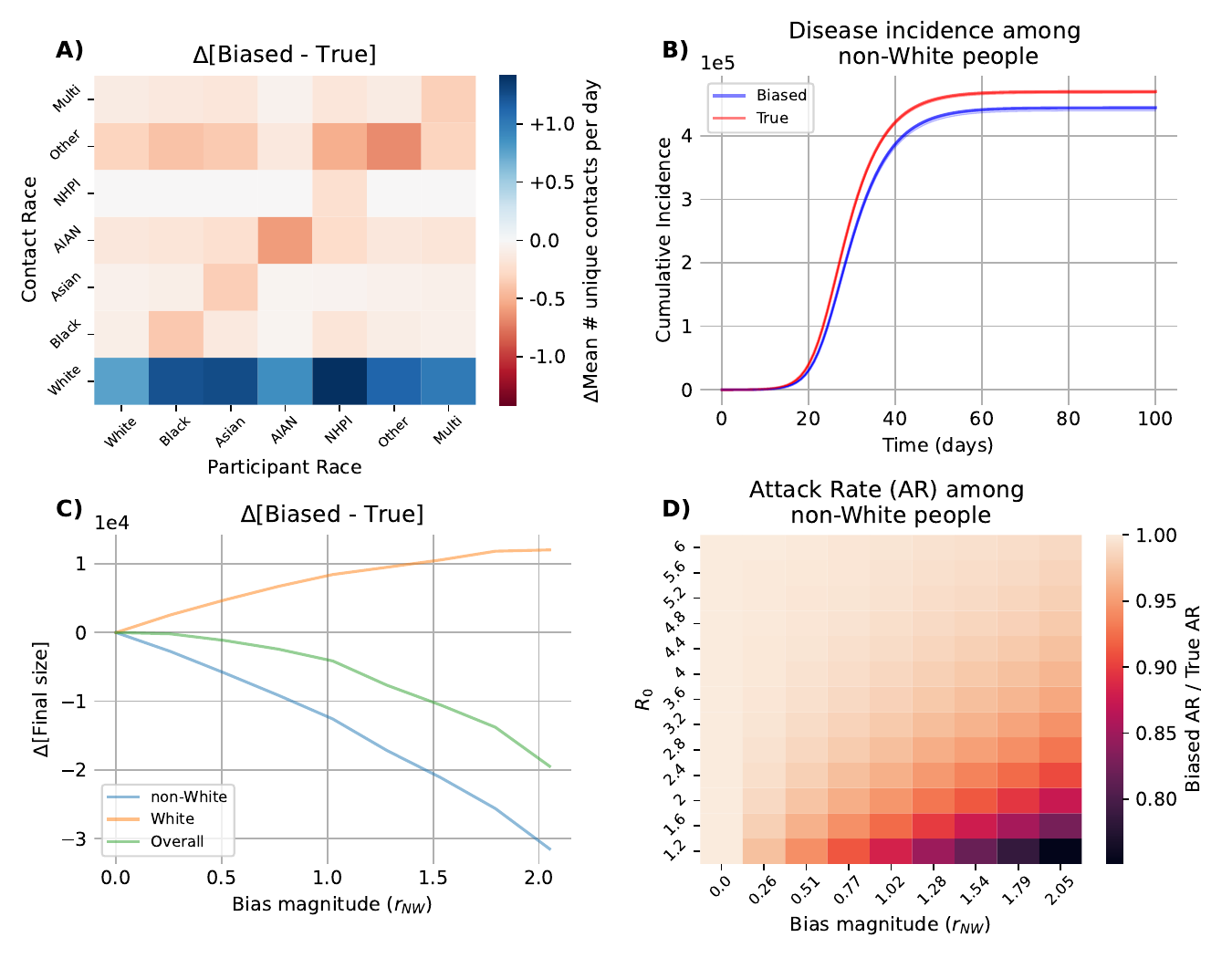}
    \caption{Impact of bias affecting the estimation of contact race ($r_\text{NW}=1.79$) on estimated contact patterns (A) and epidemic dynamics (B, C \& D), parametrised with uniform age-specific susceptibility. Racial misidentification typically occurs at a higher rate for non-White individuals when observers are presented with limited visual and audio stimulus, similar to information available in casual contact settings. By comparing the average contact matrix from ten simulated surveys with biased (`Biased') and unbiased (`True') recall, surveys affected by race perception bias lead to an underestimation of reported contact with non-White individuals individuals (A). Using an SIR model of disease spread ($R_{0}=2.9$), simulated epidemic dynamics which relied on biased survey data underestimated the true disease burden among non-White individuals (B). Increasing the magnitude of non-White racial estimation bias ($r_\text{NW}$) lead to an increasing underestimation of attack rate among non-White individuals and an increasing attack rate in the White population (C). The ratio of the simulated attack rate (AR) using the biased contact survey and the true contact survey data (`Biased AR / True AR') decreased when assuming larger $r_\text{NW}$ and a lower basic reproduction number (D).}
    \label{fig:no_age_race}
\end{figure}

\clearpage
\section{Comparison of model racial identification accuracy to experimental studies}
\label{race_empirical}
\begin{figure}[h]
    \centering
    \includegraphics[width=0.6\linewidth]{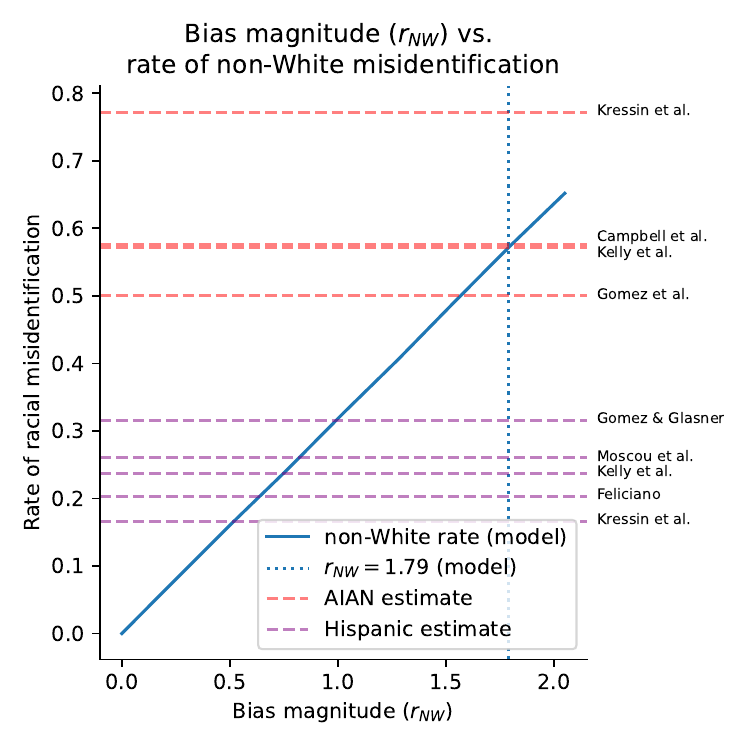}
    \caption{Comparison of model racial identification accuracy to experimental studies. Increasing the magnitude of the bias associated with identification of non-White individuals ($r_{NW}$) produces a linear increase in the rate of non-White misidentification in our model (solid blue line). Comparing the resulting rate of misidentification in our model to studies quantifying this rate from data (dashed horizontal lines) shows how realistic different rates are. The rate assumed in our main analysis ($r_{NW}=1.79$; dotted vertical line) aligns closely with the estimated rate of AIAN misidentification.}
    \label{fig:fig4_emp}
\end{figure}

To establish whether our simulation of racial bias aligned with empirical estimates, we compared the rate of racial misidentification in our model to estimates from other studies. The two main racial groups in New Mexico other than White are AIAN and Other. When using the OMB standards of racial categorisation, as assumed in this study, the Other racial group often captures Hispanic individuals. This is reflected in our synthetic population where $\sim98\%$ of individuals identifying racially as Other also identify as Hispanic. Studies characterising the rate of racial and ethnic misidentification among AIAN and Hispanic individuals have compared self-identification of race and ethnicity to an observer identification in different settings. For example, Campbell and Troyer \cite{campbell2007implications} compared racial self-identification to the racial identification recorded by interviewers in the National Longitudinal Survey of Adolescent Health. They found individuals who `self-identify as American Indian were uniquely likely to be racially misclassified by an observer' \cite{campbell2007implications}. We extracted the rate of misidentification for AIAN and Hispanic individuals from seven studies \cite{kressin2003agreement,kelly1996race,campbell2007implications,gomez2006misclassification,feliciano2016shades,moscou2003validity,gomez2005inconsistencies}. 
\\\\
We plotted the relationship between the race-related perception bias scaling factor for non-White individuals $r_{NW}$ and rate of misidentification in the community transmission setting for all non-White individuals in our model (Figure \ref{fig:fig4_emp}). We included the estimates from studies of racial misidentification as dashed horizontal lines. We limited our analysis to only community contacts as these contacts are most representative of the contexts considered in the studies of racial misidentification (e.g., hospitals). The bias scaling factor value chosen in our main analysis ($r_{NW}=1.79$) aligns most closely with the estimates of racial misidentification associated with AIAN individuals. Assuming a value for $r_{NW}$ more aligned with the Hispanic estimates (e.g., $r_{NW}\sim0.8$) reduced the bias in estimates of disease burden among non-White individuals (Figure 4C \& D).

\clearpage
\section{Transmission Setting Contribution to $\Delta$[Final Size]}
\label{trans_setting_SA}
\begin{figure}[h]
    \centering
    \includegraphics[width=\linewidth]{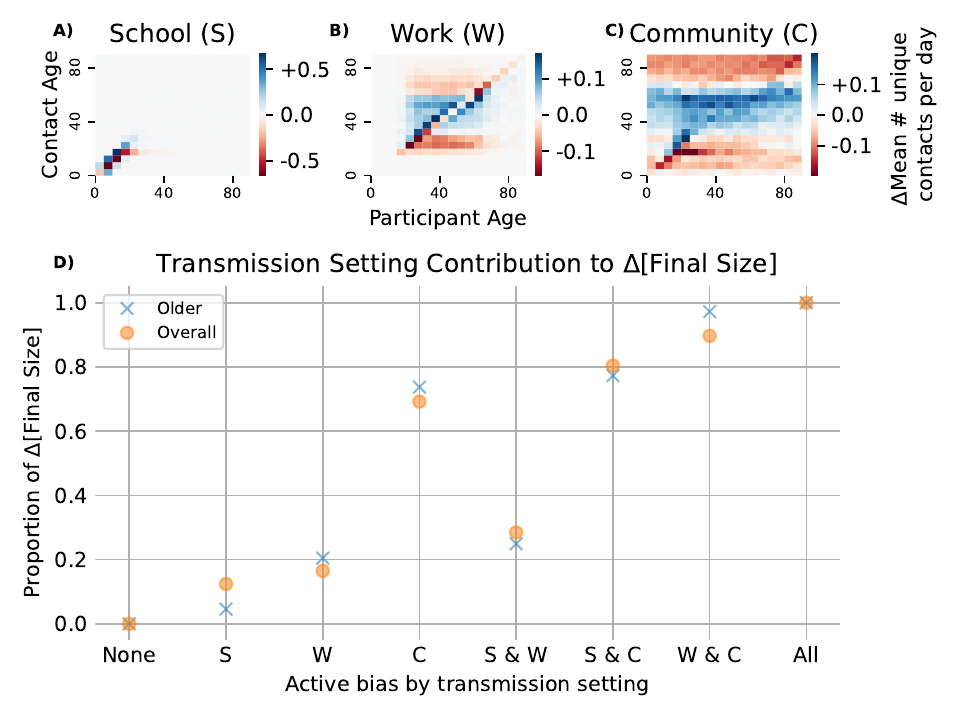}
    \caption{Transmission Setting Contribution to $\Delta$[Final Size] in age-stratified SIR model. Each transmission setting ($k$) varies in the amount of contact occurring between individuals ($c_k$), the recall accuracy of survey participants ($\epsilon_k$), and the age of individuals typically interacting in this context. Isolating the effect of bias in Community (C), School (S) and Work (W) transmission settings, by simulating contact surveys where bias in other settings is set to zero, revealed different changes in the average contact rate between age groups (panels A-C). Using these contact matrices (as well as matrices constructed from surveys with bias in two settings (i.e., C \& S, C \& W, S \& W)) to parametrise age-related contact rates in an SIR model revealed varying impacts on outbreak dynamics. We compared the change in outbreak final size ($\Delta[\text{Final Size}]$ = Final size under bias - Final size under no bias) in the whole population (overall) and the subpopulation of older people to when bias is active in all transmission settings (panel D). We found the community (C) setting made up the largest proportion of the change in outbreak final size in both the whole population and among older people.}
    \label{fig:final_size_age}
\end{figure}

\begin{figure}[h]
    \centering
    \includegraphics[width=\linewidth]{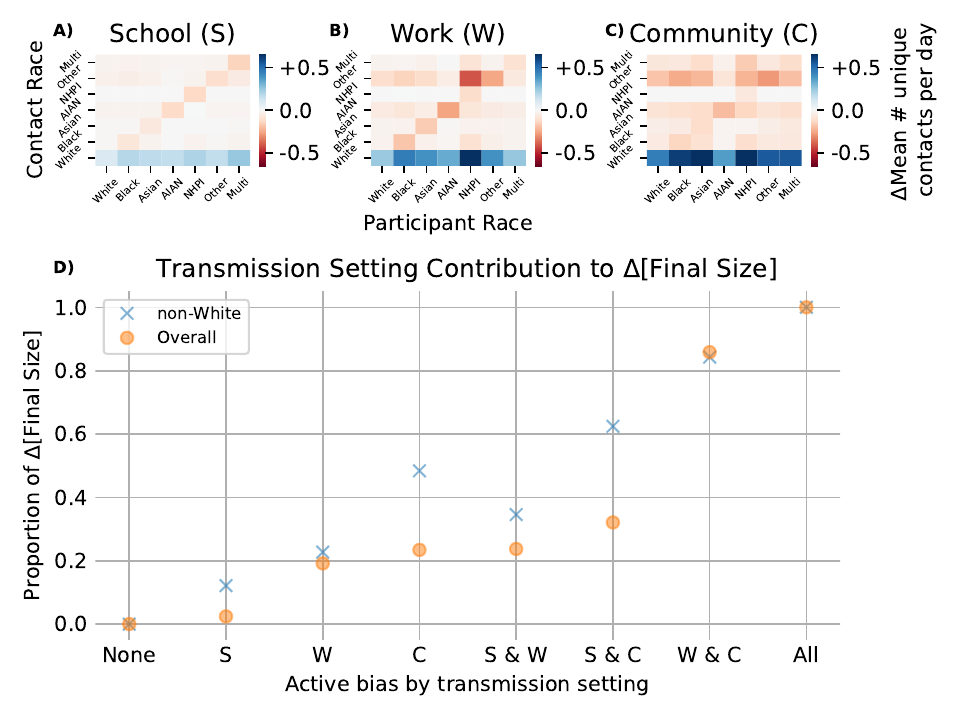}
    \caption{Transmission Setting Contribution to $\Delta$[Final Size] in age and race stratified SIR model. Each transmission setting ($k$) varies in the amount of contact occurring between individuals ($c_k$), the recall accuracy of survey participants ($\epsilon_k$), and the age of individuals typically interacting in this context. Isolating the effect of bias in Community (C), School (S) and Work (W) transmission settings, by simulating contact surveys where bias in other settings is set to zero, revealed different changes in the contact rates between racial groups (panels A-C). Using these contact matrices (as well as matrices constructed from surveys with bias in two settings (i.e., C \& S, C \& W, S \& W)) to parametrise age and race related contact rates in an SIR model revealed varying impacts on outbreak dynamics. We compared the change in outbreak final size ($\Delta[\text{Final Size}]$ = Final size under bias - Final size under no bias) in the whole population (overall) and the subpopulation of non-White people to when bias is active in all transmission settings (panel D).}
    \label{fig:final_size_race}
\end{figure}

To better understand how the perception biases introduced in our model of contact surveys impacted the simulated epidemic dynamics, we decomposed the effects of the bias on outbreak final size by transmission setting: community (C), school (S) and work (W). The household transmission setting was not considered as household contacts were assumed to be recalled without error in all simulations (i.e., $\epsilon_H=0$). We computed the mean contact matrix from ten simulated contact surveys under different assumptions of the setting-specific scaling factor ($\epsilon$). We systematically set each setting scaling factor to zero to measure each settings' effect on the final size of the outbreak (Table \ref{tab:sen_analysis_varsigma}). We used each contact matrix to simulate disease spread in a stratified SIR model. As in our main analysis, we assumed the reproduction number ($R_0=2.9$), and the age-specific susceptibility were similar to COVID-19. We assumed the age-related bias magnitude ($a$) was 2.56, and race-related bias magnitude ($r_{\text{NW}}$) was 1.79.
\begin{table}[h]
\centering
\begin{tabular}{|l|l|}
\hline
\textbf{Experiment} & \textbf{Setting-specific scaling factors ($\epsilon$)}       \\ \hline
None       & $\epsilon_S = 0$, $\epsilon_W = 0$, $\epsilon_C = 0$    \\ \hline

School (S)          & $\epsilon_S = 0.25$, $\epsilon_W = 0$, $\epsilon_C = 0$    \\ \hline
Work (W)            & $\epsilon_S = 0$, $\epsilon_W = 0.44$, $\epsilon_C = 0$    \\ \hline
Community (C)       & $\epsilon_S = 0$, $\epsilon_W = 0$, $\epsilon_C = 0.39$    \\ \hline
S \& W              & $\epsilon_S = 0.25$, $\epsilon_W = 0.44$, $\epsilon_C = 0$ \\ \hline
S \& C              & $\epsilon_S = 0.25$, $\epsilon_W = 0$, $\epsilon_C = 0.39$ \\ \hline
W \& C              & $\epsilon_S = 0$, $\epsilon_W = 0.44$, $\epsilon_C = 0.39$ \\ \hline
All              & $\epsilon_S = 0.25$, $\epsilon_W = 0.44$, $\epsilon_C = 0.39$ \\ \hline
\end{tabular}
\caption{Parameter settings for sensitivity analysis of setting-specific scaling factors ($\epsilon_k$)}
\label{tab:sen_analysis_varsigma}
\end{table}
\\\\
For each experiment $X$, we measured the change in overall and subpopulation outbreak final size, as a proportion of the total change in final size when assuming perception bias is present in all settings:
\begin{equation}
    \text{Proportion of $\Delta$[Final Size]}= \frac{\text{Final Size under $X$}-\text{Final Size under \textit{None}}}{\text{Final Size under \textit{All}}-\text{Final Size under \textit{None}}} \,.
\end{equation}
We found the change in outbreak final size observed in our simulation of age-related perception bias (Figure 3) came about mostly from bias in community contacts, and to lesser extent from workplace and school contacts (Figure \ref{fig:final_size_age}). The effect of bias in community contact settings is likely higher due to its broad impact across age groups (i.e., all age groups experience some community contact), and the higher relative contact rate in community settings (Table 1). School and workplace contact only affect specific age groups that attend these settings (Figure \ref{fig:final_size_age}A \& B). The difference in affected age groups explains the relative impact of the bias in each transmission setting on the outbreak final size among older people (Figure \ref{fig:final_size_age}D). Specifically, we can see the age of older contacts is underestimated among community contacts and to a lesser extent among workplace contacts leading to these settings accounting for a substantial proportion of the change in final size among older people. In school settings, older people are largely unaffected leading to a smaller impact on outbreak final size.
\\\\
We found the differences in outbreak final size observed in our simulation of race-related perception bias (Figure 4) were largely explained by changes in community and workplace contact patterns (Figure \ref{fig:final_size_race}). Community and workplace settings had higher relative setting-specific scaling factors and accounted for a greater proportion of overall contact when compared to the school setting. This lead to a higher magnitude reduction in reported contact with non-White individuals in community and work settings (Figure \ref{fig:final_size_race}A-C). Simulating outbreaks with perception bias affecting contact race estimation in both community and work settings (`C \& W') accounted for the majority of the change in final size in the overall population and among non-White individuals (Figure \ref{fig:final_size_race}D). Unlike the age-related bias, we observed substantial differences in the proportion of total change in final size observed between the overall and subpopulation, in this case non-White individuals. Specifically, the non-White final size changed more substantially in response to the addition of different bias settings, most notably the community contact setting in `C' and `C \& S'. This difference is consistent with the approximately linear decrease in outbreak final size among non-White, and the non-linear decrease observed in the overall outbreak final size in response to increasing bias magnitude in Figure 3. Under smaller bias ($0\leq r_{\text{NW}}\leq1$ in Figure 3), the rate of reduction in overall final size is smaller relative to a larger bias ($1.5\leq r_{\text{NW}}\leq2$ in Figure 3). When we only simulate bias in particular settings, the change in contact patterns is similar to these low bias settings where the relative change in outbreak final size is larger in the non-White population.

\clearpage
\section{Impact of model underestimation on hospital burden in New Mexico}
\label{nm_hospitals}
To estimate the proportion of COVID-19 infections among older people (65+ years) that require hospitalisation, we used data reported by the Australian Institute of Health and Welfare (AIHW) \cite{AIHW}. In this dataset, hospitalisation data is reported in 10 year age brackets. As such, we examined the age cohort for individuals aged 70+ to approximate the hospitalisation rate among older people. By dividing the total number of individuals aged 70+ hospitalised due to COVID-19 infection by the total number of individuals aged 70+ being diagnosed with COVID-19, we were able to estimate the proportion requiring hospitalisation:
\begin{equation*}
    \%\text{ Requiring hospitalisation} = \frac{31337}{377582} = 8.3\% \,.
\end{equation*}
Under our two modelling scenarios, the prevalence curve estimated using the biased contact rates underestimates the peak infections by $\sim 10,000$ infections (Figure \ref{prev}). To predict the difference in hospitalisations this corresponds to, we multiplied the number of additional infections by our estimated proportion requiring hospitalisation:
\begin{equation*}
    \#\text{ Additional hospitalisations}=0.083\times10000=830 \text{ hospitalisations} \, .
\end{equation*}
Finally, to convert the number of additional hospitalisations to a proportion of available beds, we divided by the total number of beds in New Mexico:
\begin{equation*}
    \%\text{ Percentage difference by NM hospital bed capacity}=\frac{830}{3800}=21.8\% \,.
\end{equation*}

\clearpage
\bibliographystyle{vancouver}
\bibliography{ref}